\documentclass[prx,floatfix,superscriptaddress,longbibliography,notitlepage, twocolumn, 10.5pt]{revtex4-2}

\usepackage{braket}
\usepackage{subcaption}
\usepackage{dsfont}
\usepackage{hyperref}
\usepackage[mark= ]{sectionbreak}

\usepackage{graphicx}
\graphicspath{{images/}}
\usepackage{dcolumn}
\usepackage{bm}
\usepackage{mathtools}
\usepackage{amssymb}
\usepackage[thinc]{esdiff}
\usepackage{textcomp}
\usepackage{xcolor}
\usepackage{caption}
\begin{document}

\title{ Mitigating small gaps in quantum annealing through recursive $XX-$catalyst construction }

\author{Natasha Feinstein}
\email{ucapnjs@ucl.ac.uk}
\affiliation{%
 London Centre for Nanotechnology, University College London, WC1H 0AH London, UK
}%

\author{Sebastian Schulz}
\affiliation{%
Jülich Supercomputing Centre, Institute for Advanced Simulation,
Forschungszentrum Jülich, 52425 Jülich, Germany
}%

\author{Robert J.~Banks}
\affiliation{%
 London Centre for Nanotechnology, University College London, WC1H 0AH London, UK
}%

\author{Roopayan Ghosh}
\affiliation{%
 Department of Physics and Astronomy, University College London, London WC1E 6BT, UK
}%
\affiliation{School of Basic Sciences, Indian Institute of Technology, Bhubaneswar, 752050, India}

\author{Kristel Michielsen}
\affiliation{%
Jülich Supercomputing Centre, Institute for Advanced Simulation,
Forschungszentrum Jülich, 52425 Jülich, Germany
}%
\affiliation{Universit\"at zu K\"oln, 50923 K\"oln, Germany}

\author{Sougato Bose}
\affiliation{%
 Department of Physics and Astronomy, University College London, London WC1E 6BT, UK
}%

\author{P A Warburton}
\affiliation{%
 London Centre for Nanotechnology, University College London, WC1H 0AH London, UK
}%
\affiliation{%
 Department of Electronic \& Electrical Engineering, University College London, WC1E 7JE London, UK
}%

\date{\today}


\begin{abstract}
The introduction of $XX-$interactions, typically in the form of an additional catalyst Hamiltonian, has shown promise as a tool for gap enhancement in quantum annealing, allowing the algorithm to obtain optimal solutions with shorter run-times. However, the optimal placements of these couplings within the problem graph has been found to be highly setting-dependent. In this work, we propose an iterative catalyst-construction protocol for the removal of a common quantum annealing bottleneck known as perturbative crossings. The method utilises the measurements returned by successive annealing runs to effectively steer the algorithm away from the local optima in which it becomes trapped. First, we motivate a framework for $XX-$term selection based on perturbation theory and establish its effectiveness at removing perturbative crossings on scalable toy instances with up to $39$ spins, demonstrating a consistent performance both with increasing system size and number of local optima. We then simulate the recursive catalyst-construction method on a test set of $18-$spin maximum-weight independent set problems that are generated to produce annealing spectra with multiple perturbative crossings. We find that the protocol significantly outperforms both catalyst-free annealing and annealing with catalysts containing randomly selected $XX-$terms.
\end{abstract}

\maketitle

\section{ Introduction \label{sec:introduction} }

Combinatorial optimisation encompasses a class of problems in which one is looking to either maximise or minimise an objective function over a discrete, but often exponentially large, configuration space. The real world utility of being able to solve such problems ranges from portfolio optimisation \cite{rubiogarcia2022portfolio} and resource allocation \cite{Gabbassov_2022,witt2022tactile,otsuka2022highspeed} to transport optimisation \cite{sales2023adiabatic}. In such applications, the speed with which optimal solutions can be found is often critical. Quantum annealing (QA) \cite{Apolloni1989, Finnila1994, Kadowaki1998} is a continuous time quantum algorithm that was proposed as a means of solving combinatorial optimisation problems faster than can be achieved classically by leveraging quantum tunnelling to escape local minima, which present one of the main obstacles for classical algorithms.

The run-time of QA is dependent on the energy spectrum associated with an interpolation between a trivial driver Hamiltonian (typically a homogeneous transverse-field in whose ground state a quantum system is initialised) and a non-commuting classical Hamiltonian whose ground state encodes the optimal solution to be found. For the dynamics to proceed adiabatically, such that the optimal solution is obtained at the end of the anneal, the time over which this evolution must occur scales inversely with the square of the minimum gap separating it from the first excited state \cite{Albash_2018}. Unfortunately, this gap minimum is generally expected to close exponentially with system size \cite{Smelyanskiy2001,Amin2008,Jorg2010,Young2010,Hen_2011,Seoane2012}. A particular concern is the formation of so-called perturbative crossings that are associated with the system localising in the wrong potential well \cite{Amin2009,Altshuler2010}. When this occurs, the tunnelling rate associated with reaching the ground state is exponentially suppressed in the Hamming distance between the global optimum and the local optimum in which the algorithm has become trapped. This can generally be expected to increase linearly with the system size, resulting in a tunnelling rate that is exponentially suppressed. 

The question of what kind of speed-up can be expected for quantum algorithms applied to computationally hard problems is a complex topic of discussion that lies well outside the scope of this work \cite{Bennett_1997,Abrams_1998,R_nnow_2014,Au_Yeung_2023,Preskill_2018}. In general, the aim of heuristic quantum algorithms such as QA is to introduce an enhancement to the scaling exponent for such problems, rather than change the nature of the scaling itself -- potentially identifying problems that are especially difficult for classical algorithms but where quantum approaches might be of use. However, the same properties of the local optima that introduce challenges for classical algorithms (\textit{e.g.} wider potential wells and large separation from the global optimum in the solution space while being close in solution quality) also increase the likelihood of QA becoming trapped in them and suppress the tunnelling rate associated with a transition to the global optimum \cite{Amin2008,Amin2009,Altshuler2010}. This suggests that QA is likely to struggle precisely where classical approaches do. Indeed, there is as of yet little evidence of definitive speed-up for QA over classical algorithms for the purpose of combinational optimisation (though there do exist some provable speed-ups in other contexts \cite{Somma2012,Hastings_2021,Roland_2002}). There is, therefore, an interest in developing alternative approaches to QA that produce more favourable time-scaling exponents. 

A number of potential strategies are being explored in this respect, including inhomogeneous driving \cite{Dziarmaga2010,Farhi2011,Dickson2011a,Dickson2012,Rams2016,Susa2018,Adame2020,Juh_sz_2023}, counter-diabatic driving \cite{delCampo2012counterdiabatic,Claeys2019counterdiabatic,Passarelli_2020} and the introduction of additional longitudinal terms \cite{Albash2021diag,cáliz2026reshapingquantumannealinglandscapes}. The inclusion of higher-order transverse interactions has also been proposed as a means of enhancing tunnelling rates by use of correlated spin flips. In particular, there is  large body of work demonstrating the effectiveness of $XX-$interactions for gap enhancement in QA \cite{SekiYuyaandNishimori2012,Hormozi2017,Nishimori2017,Albash2019,Mehta2021,feinstein2024,Nutricati2025,ghosh2024exponentialspeedupquantumannealing}. These interactions are typically introduced in the form of an additional \textit{catalyst} Hamiltonian which changes the path that the anneal takes between the initial and final Hamiltonian. 

The best strategy for introducing such catalysts remains unclear. Numerical studies examining a range of problem settings have found that stoquastic catalysts (\textit{i.e.}, non-complex catalysts introduced with the same sign as the driver) are generally more successful at enhancing the minimum spectral gap than their \textit{non}-stoquastic counterparts \cite{crosson2014differentstrategiesoptimizationusing,Zeng2016,Hormozi2017,Mehta2021}, a finding that has been supported by analytical arguments \cite{Crosson2020a}. On the other hand, in the settings where non-stoquastic interactions \textit{are} the more effective choice, the gap enhancements that they achieve are more significant than those reported for stoquastic catalysts \cite{crosson2014differentstrategiesoptimizationusing,Zeng2016}, with demonstrations of exponential gap enhancement for the analytically tractable $p$-spin and Hopfield models \cite{SekiYuyaandNishimori2012,Nishimori2017} as well as for specific, numerically simulated, Ising problems \cite{Albash2019,Takada2020a}. Other studies have drawn attention to the importance of the placement of $XX-$terms within a problem graph \cite{Takada2020a,Choi2021,feinstein2024,choi2026exponentialquantumspeedupstructured,ghosh2024exponentialspeedupquantumannealing}, with the placement affecting both the capacity for gap enhancement and also whether the inclusion of stoquastic or non-stoquastic couplings is more beneficial. 

Such findings motivate the idea of actively incorporating accessible, problem-specific, information into the construction of the catalyst. This could be information of the problem graph structure which is known a priori. Another option, however, would be to utilise information in the quantum states that are returned by ``failed'' annealing runs -- \textit{i.e.}, those that do not return the ground state. Repeated annealing runs are already an assumed part of the QA algorithm. Since limitations to hardware coherence times render exponentially long annealing runs impractical, it is preferable to perform an exponential number of shorter annealing runs that all achieve an exponentially small overlap with the ground state but that, taken together, result in the same probability of obtaining the global optimum as an exponentially long coherent anneal would have. The idea of recursively utilising measurements returned by the annealer to shape subsequent annealing runs has previously been explored with respect to inhomogeneous driving \cite{Dickson2011a,Dickson2012} and, more recently, as part of a method which reshapes the problem Hamiltonian itself to guide the anneal to lower energy solutions \cite{Schulz_2025}. The use of measurements has also been used as part of a catalyst-construction method to combat the appearance of small gaps for specific problem structures \cite{Choi2021}.

Following this same concept, and building on our previously published work  \cite{feinstein2024}, here we propose a general method for iteratively constructing $XX-$catalysts based on the measurements returned by successive annealing runs. Specifically, the protocol aims to mitigate the appearance of perturbative crossings by effectively guiding the anneal away from the local optima that the algorithm becomes trapped in. The theoretical motivation for the coupling selection process is not specific to a particular problem type and does not rely on specific problem graph structures, but rather leverages common features within the problem spectrum.

The structure of this paper is as follows. We begin, in section \ref{sec:theory-QA}, by defining the QA algorithm in its standard formulation and outline the theory behind the formation of perturbative crossings. Section \ref{sec:theory-removal} then uses perturbation theory to motivate specific placements of $XX-$interactions for the removal of such perturbative crossings. This expands on and solidifies the theoretical background presented in our previous work. In section \ref{sec:numerics}, we introduce a family of toy problems based on those used in \cite{feinstein2024} which provide a scalable test-bed for the catalysts that we propose. In our previous work, we demonstrated the effectiveness of the analytically motivated catalysts on systems with up to $13$ spins. Here, we utilise symmetries within the problem to obtain annealing spectra for systems with up to $39$ spins, for which we confirm that a catalyst containing a linearly scaling number of targeted $XX-$couplings is sufficient to remove a single perturbative crossing. Furthermore, we demonstrate that the performance of the catalyst is consistent with increasing numbers of perturbative crossings which the catalyst is constructed to remove. We also investigate the precision in catalyst strength needed when different numbers of couplings are included in the catalyst. 

Finally, in section \ref{sec:algorithm}, we build these ideas into a protocol for recursively constructing $XX-$catalysts based on the measurements returned by successive annealing runs. We simulate this method for a set of quasi-randomly generated instances of the maximum-weight independent set problem and find that the updating catalyst is able to guide QA to the global optimum significantly faster than both catalyst-free annealing runs and those containing randomly selected $XX-$terms. We summarise and discuss our findings in \ref{sec:conclusion}.

\section{ Perturbative crossings in quantum annealing \label{sec:theory-QA} }

Quantum annealing, in its most standard form, operates by interpolating between two non-commuting Hamiltonians. The first is a simple driver Hamiltonian, $\hat{H}_d$, and the second is the problem Hamiltonian, $\hat{H}_p$, encoding the optimisation problem to be solved and diagonal in the computation basis. The total Hamiltonian can be written as
\begin{equation}
    \hat{H}(s) = (1-s) \hat{H}_d + s \hat{H}_p,
    \label{eq:total_H}
\end{equation}
where $s$ is a dimensionless annealing time that is varied monotonically from zero to one as some function of time, $s(t)$. We refer to the eigenstates of this Hamiltonian  as the \textit{instantaneous} eigenstates and denote them and their energies as
\begin{equation}
    \hat{H}(s) \ket{E_a(s)} = E_a(s)  \ket{E_a(s)}.
    \label{eq:inst_eigenstates}
\end{equation}

For much of this work, we will be focusing on the evolution of the instantaneous eigenstates. Where the schedule does become relevant, in section \ref{sec:algorithm}, we will use the linear schedule, $s=t/T$, where $T$ is the total annealing time.

We take, as our driver, the traditional homogeneous transverse field across all the qubits,
\begin{equation}
    \hat{H}_d = - \sum \limits_{i=0}^{n-1} \hat{\sigma}^x_i,
    \label{eq:Hd}
\end{equation}
where $n$ is the total number of qubits and $\hat{\sigma}^x_i$ denotes the Pauli-X operator on the $i^\text{th}$ qubit. The problem is encoded into an Ising Hamiltonian,
\begin{equation}
    \hat{H}_p = \sum \limits_{i=0}^{n-1} h^z_i \hat{\sigma}^z_i + \sum \limits_{i=0,\text{ }j>i}^{n-1} J^{zz}_{ij} \hat{\sigma}^z_i \hat{\sigma}^z_j,
    \label{eq:Hp_ising}
\end{equation}
in such a way that the GS of $\hat{H}_p$ encodes the optimal solution to the problem. We refer to the eigenstates of $\hat{H}_p$ as the \textit{problem} eigenstates and denote them and their energies as,
\begin{equation}
    \hat{H}_p \ket{E_a} = E_a  \ket{E_a}.
    \label{eq:prob_eigenstates}
\end{equation}
Since the problem Hamiltonian is diagonal in the computational basis, the set of states $\{\ket{E_a} \}$ can be considered to be the computational basis ordered by energy in $\hat{H}_p$.

The anneal begins with $\hat{H}(s) = \hat{H}_d$ and the system initialised in the known GS. For the driver given by equation \ref{eq:Hd}, this is the symmetric superposition over all the computational basis states,
\begin{equation}
    \ket{E_0(s=0)} = \frac{1}{\sqrt{2^n}} \sum \limits_{a=0}^{2^n-1} \ket{E_a}.
    \label{eq:Hd_GS}
\end{equation}
The aim is then to interpolate from $s=0$ to $s=1$ in such a way that the evolution of the system, $\ket{\Psi(s)}$, proceeds adiabatically (\textit{i.e.}, $\ket{\Psi(s)} \approx \ket{E_0(s)}$ $\forall$ $s$). If this is achieved, the evolution will end with $\ket{\Psi(s)}$ having a high overlap with the GS of $\hat{H}_p$ such that measuring the system returns the optimal solution with high fidelity. The rate at which the evolution can progress while maintaining adiabaticity is described by the adiabatic theorem. This theorem has a number of formulations (see \cite{Albash_2018} and references within) but in its most basic form it states that the total time for which the anneal is run must scale inversely with the square of the minimum spectral gap separating the ground and first excited state,
\begin{equation}
    T \propto \frac{1}{\min \limits_{0 \leq s \leq 1} [E_1(s) - E_0(s)]^2  }.
    \label{eq:adiabatic}
\end{equation}
If this condition is not met, GS amplitude may be lost to higher excited states as the system passes through the locations of small spectral gaps. 

A key take-away from equation \ref{eq:adiabatic} is that, if the minimum spectral gap in the instantaneous annealing spectrum is expected to close exponentially with the problem size, there will be a corresponding exponential \textit{increase} in the annealing time required to obtain the GS of $\hat{H}_p$ with high fidelity. Unfortunately, exponentially closing gaps are an expected feature in the quantum annealing of hard problems \cite{Amin2008,Young2010,Hen_2011}. 

A particular problem is the formation of so-called perturbative crossings (PCs) which can form between competing local optima \cite{Amin2009,Altshuler2010}. Their formation can be understood by introducing $\hat{H}_d$ as a perturbation to the problem Hamiltonian and determining whether the energy of any of the perturbed excited states crosses with that of the perturbed problem ground state (GS) \cite{Amin2009,Altshuler2010}. If such a crossing is predicted by perturbation theory then we can expect the annealing spectrum to exhibit an avoided level crossing (AC) at the corresponding point in the anneal and it is this feature that we refer to as a perturbative crossing.

Carrying out this analysis for specific problem instances of interest is of course impractical as it requires knowledge of the problem GS that necessitates the problem already being solved. Rather, we can examine the perturbative corrections in their most general form to help us understand how and when we expect these PCs to occur. The majority of the analysis in this section has been carried out in some form across different works by other authors \cite{Amin2009,Altshuler2010,Dickson2011a,Choi2020} -- though often for specific systems. Our aim here is to consolidate this analysis for a general setting and make explicit the key points relevant to this work.

\subsection{Formation of a perturbative crossing \label{sec:theory-PCs_formation} }

Given the perturbed Hamiltonian,
\begin{equation}
    \hat{H}(\lambda) = \hat{H}_p + \lambda \hat{H}_d,
    \label{eq:H-pert}
\end{equation}
the perturbed problem eigenstates are given, to second order in $\lambda$, by
\begin{multline}
    E_a(\lambda) = E_a + \lambda \braket{E_a|\hat{H}_d|E_a} + \lambda^2 \sum \limits_{b \neq a} \frac{|\braket{E_b|\hat{H}_d|E_a}|^2}{E_a - E_b} \\ = E_a + \lambda^2 \sum \limits_{b \in N_x(a)} \frac{1}{E_a - E_b},
    \label{eq:general-pert}
\end{multline}
 where we have used $N_x(a)$ to denote the set of states to which $a$ is coupled by the driver. The first order term vanishes since $\hat{H}_d$ cannot couple a problem state to itself and the second order term is simplified by the fact that $|\braket{E_b|\hat{H}_d|E_a}|=1$ for $b \in N_x(a)$ and $0$ otherwise. 

Two perturbed states, with indices $a$ and $b$ ($E_b(0) > E_a(0)$), will cross if there exists some $\lambda_\text{x}$ for which $E_b(\lambda) \leq E_a(\lambda)$ for $\lambda \geq \lambda_\text{x}$. If this $\lambda_\text{x}$ is within the convergence radius of the expansion, this indicates the formation of an AC at some $s_\text{x}$ towards the end of the annealing spectrum. We can obtain an estimate for the location of the AC with the transformation $s=1/(1+\lambda)$ (which transforms the perturbed Hamiltonian given in equation \ref{eq:H-pert} into the annealing Hamiltonian given in equation \ref{eq:total_H}).

We are interested in crossings involving the perturbed problem GS since these are responsible for the ACs that will bottleneck the annealing algorithm. We can therefore restrict ourselves to considering perturbations to the problem GS and the problem states, $\ket{E_a}$, with $E_0 < E_a \leq E_0 + \delta$ where $\delta$ is small compared to the energy gap between the problem ground state and the rest of the spectrum -- \textit{i.e.}, states that are close enough in energy to the GS that small perturbations could cause their energies to cross. We refer to this set of states as $S_\delta$.

Let us consider a state, $\ket{E_a}$, in $S_\delta$ that corresponds to a local optimum of the problem. As we shall see, these are the states of primary concern with regard to the formation of PCs. By definition, all the states that are one spin flip away from $\ket{E_a}$ (that is, all the states in $N_x(a)$) will have energies greater than $E_a$. As such, each term in the sum from equation \ref{eq:general-pert} will be negative:
\begin{equation}
    E_a(\lambda) = E_a - \lambda^2 \sum \limits_{b \in N_x(a)} \frac{1}{|E_b - E_a|}.
    \label{eq:local_opt_perts}
\end{equation}
Technically, taking the absolute value in the denominator is redundant since $(E_b-E_a)$ will always be positive in this context. We include it, however, to make the sign of these terms explicit. Since the terms all enter the sum with the same sign, the effects of the perturbations will be cumulative. Thus, the local optima that are coupled by the driver to lower energy states will receive the greatest perturbations. 

From these arguments, we conclude that a local optimum $\ket{E_a} \in S_\delta$ will result in a PC with the ground state if it is coupled, by the driver, to more low energy states than the problem ground state is. We can think of this as the quantum annealing algorithm being more likely to get caught in wider potential wells. Note that a state in $S_\delta$ that is \textit{not} a local optimum (\textit{i.e.}, a state that is within a deep potential well but not at the bottom of it) will have both positive and negative contributions to the sum in equation \ref{eq:general-pert}. As such, it can be expected to receive a less negative overall perturbation from the driver compared to the local optimum at the bottom of the well.

Now that we have established when and how PCs form in the annealing spectrum, let us examine the size of spectral gap minima that form at the point of these crossings.

\subsection{Spectral gap at a perturbative crossing}
 
That there is a finite gap size at the PC is a result of tunnelling between the associated states. As such, the gap at a crossing between two perturbed states is proportional to their overlap. By writing out the perturbed problem states in terms of the computational basis states, we obtain 
\begin{multline}
    \Delta E_{ab}(\lambda_\text{x}) \propto \braket{E_a(\lambda_\text{x})|E_b(\lambda_\text{x})} \\ = 
     \left( \sum \limits_{c=0}^{2^n-1} c_{ac}(\lambda_\text{x}) \bra{E_c} \right) \left( \sum \limits_{d=0}^{2^n-1} c_{bd}(\lambda_\text{x}) \ket{E_d} \right)
    \\ = \sum \limits_{c = 0}^{2^n-1} c_{ac}(\lambda_\text{x})c_{bc}(\lambda_\text{x}),
    \label{eq:gap-prop}
\end{multline}
where $c_{ac}(\lambda) = \braket{E_a(\lambda)|E_c}$. In this setting $\braket{E_a(\lambda)|E_c}$ is equal to $\braket{E_c|E_a(\lambda)}$ since our Hamiltonians are real and symmetric.  

Perturbation theory tells us that, for any perturbed problem state $\ket{E_a}$, the magnitude of $c_{ac}(\lambda_\text{x})$ will be exponentially small in the Hamming distance between $\ket{E_a}$ and $\ket{E_c}$ -- since $\ket{E_c}$ will only enter the perturbative expansion of $\ket{E_a}$ on the order of this Hamming distance. Therefore, the gap minimum between two crossing states, with labels $a$ and $b$, will be exponentially small in the Hamming distance between $\ket{E_a}$ and $\ket{E_b}$. Since the average Hamming distance between states will increase linearly with the system size, the gap minima associated with PCs are expected to close exponentially with that size. 

Another consequence of equation \ref{eq:gap-prop} is that perturbative crossings occurring further from the end of the anneal will tend to have larger associated gap minima. This is because the states involved will have undergone greater perturbation prior to the point of their crossing, such that they are more mixed and are therefore likely to have a larger overlap with each other.

Having now established the conditions under which PCs will form and why they present a quantum annealing bottleneck, we are in a position to consider how the inclusion of an $XX-$catalyst could help to mitigate these conditions. 

\section{ Targeted $XX-$couplings \label{sec:theory-removal} }

We are interested in understanding how the inclusion of different $XX-$couplings in a catalyst Hamiltonian will impact the formation of PCs in the annealing spectrum. We introduce our catalyst Hamiltonian to the standard annealing Hamiltonian presented in equation \ref{eq:total_H} as follows:
\begin{equation}
    \hat{H}(s) = (1-s) \hat{H}_d + s(1-s) \hat{H}_c + s \hat{H}_p.
    \label{eq:total_H_cat}
\end{equation}
Its prefactor means that the catalyst contributes to the anneal neither at $s=0$ nor $s=1$, such that it does not interfere with initialisation or readout. We consider catalysts of the form
\begin{equation}
    \hat{H}_c = \sum \limits_{(i,j) \in E_\text{cat}} J^{xx}_{i,j} \sigma^x_i \sigma^x_j,
    \label{eq:Hc}
\end{equation}
where $E_\text{cat}$ denotes the set of edges included in the catalyst. This needs not be the same as the set off edges included in the problem graph. For the purpose of our goals in this section, we set $J^{xx}_{i,j} = J_{xx}$ for all $(i,j) \in E_\text{cat}$. Note that if $J_{xx}$ is chosen to be negative the total Hamiltonian will be stoquastic whereas if $J_{xx}$ is chosen to be \textit{positive}, $\hat{H}(s)$ will be \textit{non}-stoquastic for $0<s<1$. It has been suggested that non-stoquastic QA may be more powerful than its stoquastic counterpart as it is harder to simulate classically \cite{gupta2019elucidating} and makes adiabatic QA universal \cite{Crosson2020}. 

Let us consider what impact we would want a catalyst to have for the removal of a PC. Intuitively, we want to pull the two perturbed problem states involved in the crossing apart such that the crossing is removed. Recalling that we are interested in crossings involving the GS, this means that we would like to \textit{reduce} the energy of the perturbed problem GS and/or \textit{increase} the energy of the perturbed state that crosses it.

To understand how different choices of $E_\text{cat}$ impact the perturbed problem eigenstates, we will introduce $\hat{H}_c$ as an additional perturbation to equation \ref{eq:H-pert} and use our understanding of the make-up of the driver-perturbed states, $\ket{E_a(\lambda)}$ to deduce the effect of the catalyst on the energies. Something similar to this is carried out in \cite{Choi2021} for the specific setting considered there, though the conclusions we reach in this section hinge on different theoretical arguments. 

The twice-perturbed Hamiltonian can be written as follows:
\begin{equation}
    \hat{H}(\lambda,\mu) = \hat{H}(\lambda) + \mu \hat{H}_c.
\end{equation}
The perturbed energies are given, to first order, by
\begin{multline}
    E_a(\lambda,\mu) = E_a(\lambda) + \mu \bra{E_a(\lambda)}\hat{H}_c\ket{E_a(\lambda)} \\
    = E_a(\lambda) + \mu \left( \sum \limits_{b=0}^{2^n -1} \braket{E_a(\lambda)|E_b} \bra{E_b} \right) \hat{H}_c  \\ \left( \sum \limits_{c=0}^{2^n -1} \ket{E_c} \braket{E_c|E_a(\lambda)} \right) \\
     = E_a(\lambda) + 2 \mu J_{xx} \sum \limits_{(b,c) \in C_{xx}} c_{ab}(\lambda)c_{ac}(\lambda),
     \label{eq:non-stoq-pert}
\end{multline}
where we have used $C_{xx}$ to denote the set of problem state pairs that are coupled by the catalyst. Note that every coupling in $E_\text{cat}$ will pair up the problem eigenstates such that the number of state pairs contained in $C_{xx}$ is $|E_\text{cat}|\times 2^{(n-1)}$.

Equation \ref{eq:non-stoq-pert} tells us that, at least to lowest order in $\mu$, the effect of the catalyst on the driver-perturbed state, $\ket{E_a(\lambda)}$  is dependent on the presence of the pairs of catalyst-coupled states, $(b,c) \in C_{xx}$, in $\ket{E_a(\lambda)}$. To understand what this implies, we need to understand the make-up of the driver-perturbed states, 
\begin{equation}
    \ket{E_a(\lambda)} = \sum \limits_{b=0}^{2^n-1} c_{ab}(\lambda) \ket{E_b}.
\end{equation}

\subsection{Properties of the driver-perturbed states \label{sec:theory-state_makeup} }

Note that at this point we are considering perturbations to the \textit{problem states} from the introduction of the \textit{driver} (such that we can understand the corrections to the energies of these perturbed states as a result of a separate perturbation from the \textit{catalyst}). Perturbation theory tells us that $\ket{E_a(\lambda)}$ will be dominated by $\ket{E_a}$ and problem states, $\ket{E_b}$, that are close to $\ket{E_a}$ in Hamming distance and energy. As such, we expect larger terms to enter the sum in equation \ref{eq:non-stoq-pert} if $XX-$terms are chosen that couple $\ket{E_a}$ to states that are close in energy to it -- as well as coupling the states that are close in energy and Hamming distance to $\ket{E_a}$ to each other. 

Without knowing the signs of the overlaps, $c_{ab}$ and $c_{ac}$, that enter equation \ref{eq:non-stoq-pert}, the overall impact of the catalyst is difficult to ascertain. Fortunately, we are able to show that, for the driver-perturbed states of interest, the dominant contributions to $\ket{E_a(\lambda)}$ are positive. 

To do this, we take a more detailed look at the perturbative corrections to the problem states from the driver. Perturbative corrections of different orders introduce problem states to $\ket{E_a(\lambda)}$ with different Hamming distances from $\ket{E_a}$. Specifically, the $k^\text{th}$ order correction will introduce states that are a Hamming distance $k$ from $\ket{E_a}$ as well as some states with a Hamming distance less than $k$. 

The equation associated with the $k^\text{th}$ order correction is
\begin{equation}
    (E_a - \hat{H}_p)\ket{E_a^{(k)}} = \hat{H}_d \ket{E_a^{(k-1)}} - \sum \limits^{k}_{i=2} E_a^{(i)} \ket{E_a^{(k-i)}}.
\end{equation}
The sum starts at $i=2$ and not $1$ because, in our setting, the first order correction to the energies is zero. Pre-multiplying by some state $\ket{E_b}$ ($b \neq a$) and rearranging we obtain
\begin{multline}
    c^{(k)}_{ab}= \braket{E_b|E_a^{(k)}} = \\ \frac{\braket{E_b|\hat{H}_d|E_a^{(k-1)}}}{(E_a - E_b)} - \sum \limits^{k}_{i=2} E_a^{(i)} \frac{\braket{E_b|E_a^{(k-i)}}}{(E_a - E_b)},
    \label{eq:nth_order_pert}
\end{multline}
where $c^{(k)}_{ab}$ is the $k^\text{th}$ order contribution to $c_{ab}$. For the lowest order non-zero $c^{(k)}_{ab}$ associated with the state $\ket{E_b}$, the second term in equation \ref{eq:nth_order_pert} vanishes. 

Starting with
\begin{equation*}
    \ket{E_a^{(1)}} =  \sum \limits_{b \neq a} \frac{\braket{E_b|\hat{H}_d|E_a}}{(E_a - E_b)} \ket{E_b},
\end{equation*}
we can recursively determine that the first term in equation \ref{eq:nth_order_pert} will be a sum of terms with:
\begin{itemize}
    \item Numerators that are a product of $k$ matrix elements, $\braket{E_d|\hat{H}_d|E_c}$ where the states, $\ket{E_c}$ and $\ket{E_d}$, can be the problem state being perturbed, $\ket{E_a}$, the state we are looking for the contribution from, $\ket{E_b}$, or any other problem state.
    \item Denominators that are products of $k$ energy differences, $(E_a - E_c)$.
\end{itemize}
The matrix elements appearing in the numerators will all be either $-1$ or zero depending on whether or not the two problem states are one spin flip apart. With regard to the denominators, recall that we are interested in corrections to states, $\ket{E_a}$, that correspond to local optima -- \textit{i.e.} states that are at the bottom of potential wells. The terms, $(E_a - E_c)$, are therefore expected to be negative such that the overall sign of each of the terms will be positive. 

For this to not be the case, we would need to be considering states that are far enough from $\ket{E_a}$ to be outside of $\ket{E_a}$'s potential well and towards the bottom of another, deeper, potential well. These states are not likely to be the dominant contributors to $\ket{E_a(\lambda)}$ since they will only enter at larger orders of perturbation theory. If the dominant contributions, $c_{ab}(\lambda) = \braket{E_b|E_a(\lambda)}$, to a perturbed local optimum, $\ket{E_a(\lambda}$ ($a \in S_\delta$), are positive, the largest terms that appear in the sum in equation \ref{eq:non-stoq-pert} will also be positive. As such, the sign of $J_{xx}$ dictates whether the catalyst will raise or lower the energy of one of these perturbed states.  

Let us now use this to understand what impact different $XX-$coupling choices will have on the perturbed problem states -- and thus on a PC present in the spectrum. 

\subsection{Selecting $XX-$terms \label{sec:XX_selection} }

\begin{figure}[]
    \centering
    \includegraphics[scale=0.3]{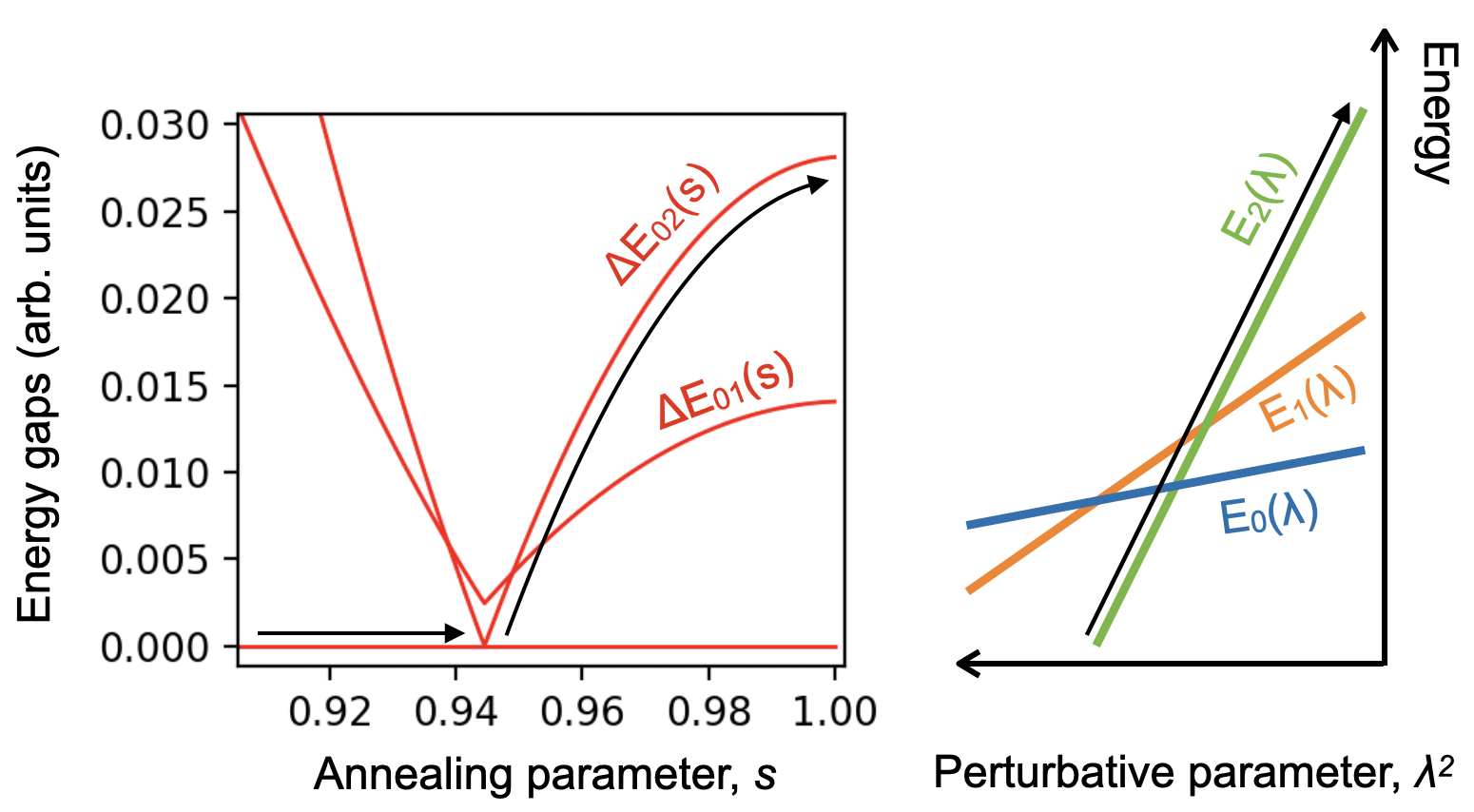}
    \caption{ Figure depicting an example perturbative crossing structure. The left plot shows the end portion of a calculated $9$-spin annealing spectrum for a problem with the parameters $n_0=2$, $n_1=3$, $n_2=4$, $W_0=1.00$, $W_1=0.99$, $W_2=0.98$ and $J_{zz}=5$ -- see section \ref{sec:numerics-setting}. The cartoon on the right illustrates the perturbed problem states associated with this spectrum. A black arrow shows, on both the spectrum and the cartoon, the expected path which an anneal would take if transitions occur at the locations of the perturbative crossings. }
\label{fig:spectrum_example}
\end{figure}

So far we have established the following: 
\begin{enumerate}
    \item The catalyst will have the greatest effect on the perturbed energy, $E_a(\lambda)$, if the $XX-$terms are chosen to couple $\ket{E_a}$ to other low energy states and/or couple together low energy states that are close in Hamming distance to $\ket{E_a}$.
    \item The energy will \textit{increase} if the catalyst is introduced with $J_{xx}>0$ - \textit{i.e.}, the \textit{opposite} sign to the driver (\textit{non-stoquastic} total $\hat{H}(s)$).
    \item The energy will \textit{decrease} if the catalyst is introduced with $J_{xx}<0$ - \textit{i.e.}, the \textit{same} sign as the driver (\textit{stoquastic} total $\hat{H}(s)$).
\end{enumerate}
These results suggest two potential routes for removing a PC that forms between two problem states, $\ket{E_a}$ and $\ket{E_b}$ ($b>a$). We could introduce couplings with $J_{xx}<0$ that target $\ket{E_a}$ to reduce $E_a(\lambda)$. Alternatively we could introduce couplings with $J_{xx}>0$ that target $\ket{E_b}$ to increase $E_b(\lambda)$. 

As already noted, we are concerned with the case where $a=0$, \textit{i.e.} where there is a crossing with the problem GS. An approach which requires us to select $XX-$terms based on information of the problem GS is likely not practical as it assumes prior access to knowledge of the global optimum. While it is not inconceivable that a setting may exist in which one has sufficient knowledge of the global optimum to select suitable couplings while still not knowing the optimal solution in full, we will focus our attention on the second approach proposed which involves targeting the \textit{local} optimum responsible for the crossing. 

This approach is especially promising since, if the PCs are the only exponentially closing gaps in the spectrum, a polynomial-time anneal is likely to return precisely the local optimum responsible for the PC. An example spectrum illustrating this is shown in figure \ref{fig:spectrum_example}. The left plot shows the end portion of a calculated annealing spectrum and on the right we show a cartoon illustrating the perturbations to the problem states that produce this spectrum. (The construction of the problem Hamiltonian that produces this spectrum will be discussed in section \ref{sec:numerics-setting}.) The black arrows illustrate the path an anneal would follow if transitions occur at the locations of the PCs. 

Having motivated the form of catalyst which we will explore, we will now present numerical results that demonstrate the effectiveness of these coupling choices. 

\section{ Demonstration of catalyst performance on toy problems \label{sec:numerics} }

In this section, we utilise a family of toy problems to demonstrate that the proposed catalyst is indeed able to enhance the gap minimum at a PC by targeting the associated local minimum. We begin, in section \ref{sec:numerics-setting}, by describing the structured MWIS instances used for this investigation. These instances allow us to controllably produce annealing spectra with different structures of PCs. Furthermore, we are able to make use of symmetries in the problems to reduce the size of the relevant Hilbert space, thereby reducing the computational complexity of our simulations and allowing us to access larger system sizes. 

In section \ref{sec:numerics-1PC} we investigate the effect of the proposed catalyst on a single PC. We examine its effect on the gap scaling when different numbers of couplings are included in the catalyst and find that including a linearly scaling number of couplings is sufficient to fully remove the PCs. Not only this, but we find that the required catalyst strength, $J_{xx}$, decreases with the system size. We also examine the precision with which the catalyst strength must be chosen in order to achieve this, finding that including more couplings in the catalyst results in greater stability of the gap enhancement to variation in $J_{xx}$. Finally, in section \ref{sec:numerics-multiple}, we demonstrate that the catalyst is able to remove multiple PCs from the same spectrum with a component targeting each of the necessary local optima. 

\subsection{ Problem setting \label{sec:numerics-setting} }

Our goal is to define a family of problem instances that allows us to control the PCs that appear in the annealing spectrum. As we shall see, the maximum-weight independent set (MWIS) problem has properties that facilitate this. We begin, in section \ref{sec:MWIS}, by outlining the MWIS problem before going on, in section \ref{sec:structured_MWIS}, to describe the family of problem instances we construct for our purposes. Finally we present some numerical \textit{catalyst free} results in section \ref{sec:structured_MWIS_catfree} to illustrate the relationship between the problem parameters and the resultant annealing spectra. 

\subsubsection{MWIS problems \label{sec:MWIS} }

The MWIS problem takes as its input an undirected, weighted graph and aims to find the set of vertices with the largest weight for which no two vertices in the set are connected by an edge. The problem is easily mapped to an Ising Hamiltonian by representing each vertex with a spin such that each eigenstate of $\hat{H}_p$ represents a set containing the vertices to which the up-spins in the state correspond. For instance, the problem state $\ket{E_a} = \ket{\downarrow \uparrow \uparrow \downarrow \uparrow}$ corresponds to the set of vertices $\{2,3,5\}$. A crucial point that will come into play later is that, by this encoding, flipping a spin corresponds to either adding or removing a vertex from the set. 

Loosely speaking, local $Z-$fields implement the vertex weights and the independent set condition is achieved through the anti-ferromagnetic $ZZ-$coupling of qubits corresponding to vertices connected by an edge in the problem graph. More specifically, the problem Hamiltonian is given by 
\begin{equation}
    \hat{H}_p = \sum \limits_{i \in \{\textrm{\scriptsize vertices}\}} (d_i J_{zz} - 2w_i)\hat{\sigma}^z_i +  J_{zz} \sum \limits_{(i,j) \in \{\textrm{\scriptsize edges}\}} \hat{\sigma}^z_i \hat{\sigma}^z_j ,
    \label{eq:Hp}
\end{equation}
where $d_i$ is the degree and $w_i$ the weight of vertex $i$. That the $ZZ-$coupling pre-factor, $J_{zz}$, appears in the local field terms is to account for the fact that two down spins coupled by an edge should not be penalised. \textit{I.e.}, it is acceptable for two adjacent spins to not be included in the set. 

For sufficiently large $J_{zz}$, the eigenspectrum of $\hat{H}_p$ consists of clusters of states whose energies depend on the number of edge dependencies present in the corresponding set of vertices. The higher the magnitude of $J_{zz}$, the greater the separation between these clusters. Within the clusters, the energies of the states will be ordered by the total weight of the corresponding set, with the highest-weighted set having the lowest energy.

\subsubsection{Structured MWIS instances \label{sec:structured_MWIS} }

Let us now outline the family of structured MWIS instances which we will be using in our simulations. From our discussion in section \ref{sec:theory-PCs_formation}, we know that controlling the formation of different PCs requires us to be able to both
\begin{itemize}
    \item set the number of local optima in the problem and how close the corresponding problem state energies are to the ground state; and
    \item adjust the energies of the states close in Hamming distance to these local optima -- \textit{i.e.}, adjust the width of the potential wells. 
\end{itemize}

To achieve this, we utilise the same kind of complete $k$-partite graphs as we have in previous works \cite{feinstein2024,feinstein2024robustnessdiabaticenhancementquantum}, considering a structure consisting of $k$ sub-graphs. Each sub-graph, $G_a$, contains $n_a$ vertices. There are no edges between vertices within the same sub-graph and all vertices within one sub-graph are connected to all the vertices within all the other sub-graphs. We allocate a total weight, $W_a$, to each sub-graph which is split evenly between the vertices such that each vertex $i \in G_a$ has a weight $w_i = W_a/n_a$. The MWIS problem on such a graph has $k$ maximally independent sets, each corresponding to selecting all the vertices in one of the sub-graphs and none from any of the others. Thus the problem will have $k-1$ local optima in addition to the global optimum.

The sub-graph weights, $W_a$, will decide which of the $k$ maximally independent sets is the global optimum and the ordering of the other local optima. Without loss of generality we can label our sub-graphs by decreasing weight ($W_0 > W_1 > W_2 \dots$) such that $G_0$ corresponds to the global optimum and $G_a$ corresponds to the $a^\text{th}$ highest-weighted local optimum. By selecting values for $W_a$ that are close together, we can ensure the problem has a set of local optima that are highly competitive with the global optimum. Correspondingly, the problem spectrum will contain $k-1$ excited states corresponding to these local optima that are close in energy to the ground state. 

The neighbourhood around the global and $k-1$ local optima can be adjusted through the sub-graph sizes, $n_a$. Recall that, for the Ising encoding of the MWIS problem, flipping a spin from down to up corresponds to adding a vertex to the corresponding set and that flipping a spin from up to down corresponds to removing a vertex. This means that a state representing a maximally independent set containing $n_a$ vertices will be coupled, by the driver, to $n_a$ states representing independent sets -- \textit{i.e.}, the states where one vertex from the set has been removed. (Correspondingly, the state will be coupled to the $\sum_{b \neq a}n_b$ states representing dependent sets where one vertex has been added to the set.) By the encoding of the problem, the states corresponding to independent sets will have significantly lower energy than those corresponding to dependent sets. Thus, $n_a$ can be used to adjust the strength of the negative perturbation from the driver to the state associated with $G_a$ -- with larger $n_a$ resulting in a stronger perturbation. 

Note that the set of sub-graph sizes, $\{n_a\}$, and weights, $\{W_a\}$, fully define the optimisation problem and that $J_{zz}$ is a parameter associated only with the Ising encoding of the problem. Note also that the solution to the MWIS problem would be unchanged if we were to scale up all the weights, $\{W_a\}$, by the same factor. However, if these altered values were inserted into equation \ref{eq:Hp}, the annealing spectrum would be significantly altered. All other factors being equal, scaling up the coupling and field strengths that enter $\hat{H}_p$ will result in gap enhancement. We therefore wish to set a consistent energy scale for $H_p$ for a given system size, $n = \sum_{a=0}^{k-1}n_a$ and a consistent relationship between the energy scales of $\hat{H}_p$ and $\hat{H}_d$ with increasing $n$. To achieve this, we normalise the parameters that enter equation \ref{eq:Hp} such that the energy scale ($E_{n-1}-E_0$) of $H_p$ is consistently ten times that of $H_d$.

Throughout this work, we will quote the \textit{un-normalised} values of $\{W_a\}$ and $J_{zz}$ as these allow for easier comprehension of the problem set-up. 

\subsubsection{ Catalyst free results \label{sec:structured_MWIS_catfree} }

\begin{figure*}[]
    \centering
    \includegraphics[scale=0.42]{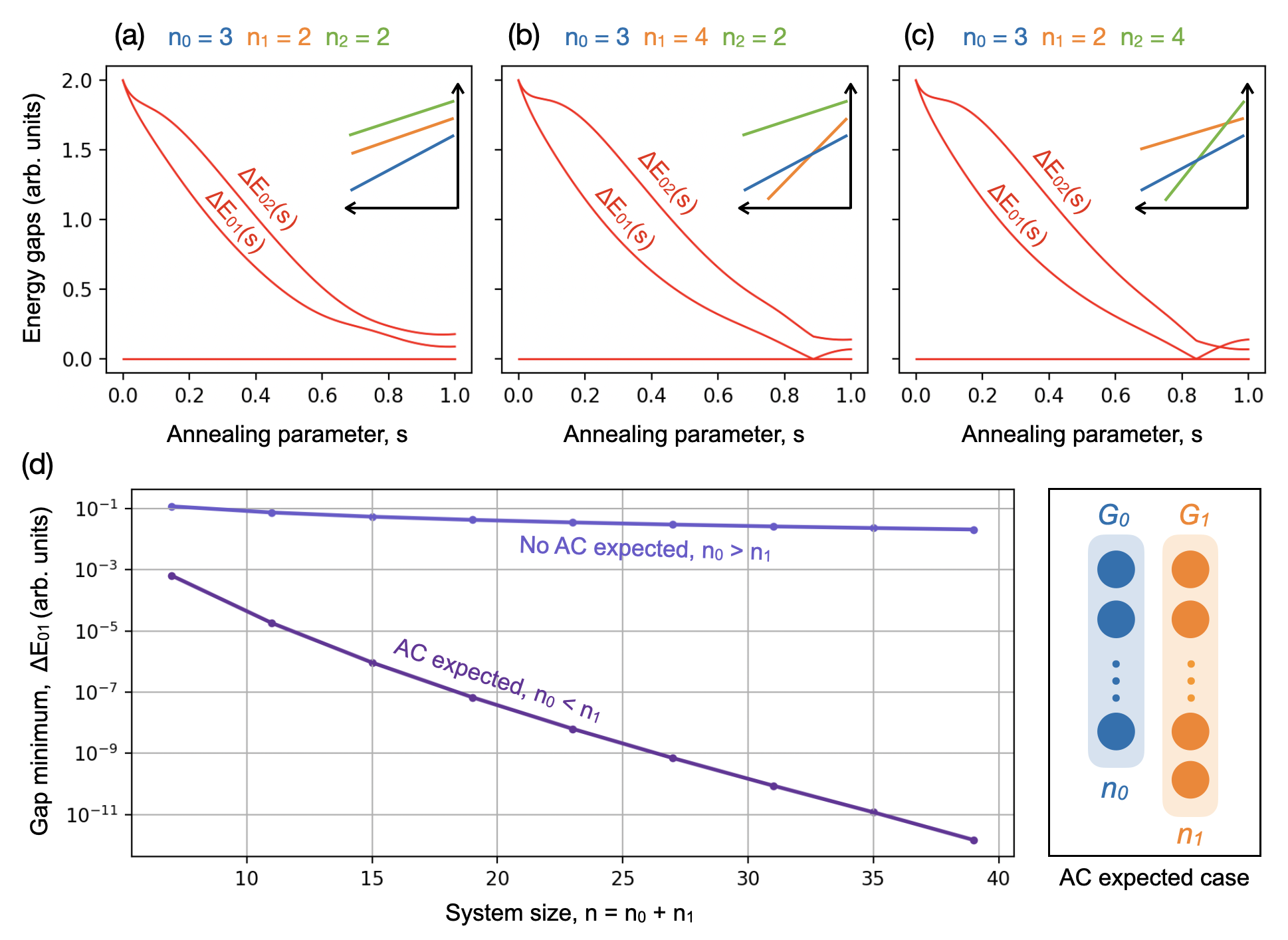}
    \caption{ Figures (a-c) show the numerically obtained gap spectra associated with three different tri-partite problem instances, constructed as described in section \ref{sec:structured_MWIS}, with $W_0=1.05$, $W_1=1.00$, $W_2=0.95$ and $J_{zz}=5$. The sub-graph sizes, $n_a$, are given above each plot and we include a cartoon illustrating the expected perturbations to the problem states associated with $G_0$ (blue), $G_1$ (orange) and $G_2$ (green). Figure (d) shows the scaling of the gap minimum for a bi-partite graph with $W_0=1.05$, $W_1=1.00$ and $J_{zz}=5$. The darker purple shows the case where $n_0 = n_1 - 1$ where we expect an avoided level-crossing (AC) to form and the lighter purple shows the case where $n_0 = n_1 + 1$ where we do \textit{not} expect an AC to form.  }
\label{fig:crossing_exampes}
\end{figure*}

We now present numerical results demonstrating that we are able to use this family of graphs to produce the desired features in the associated catalyst-free annealing spectra. Our numerical results are obtained by using closed-system spin models where we have utilised the symmetries present in the systems to reduce the size of the relevant Hilbert space -- details on how this was achieved can be found in the appendices of our previous work \cite{feinstein2024robustnessdiabaticenhancementquantum}. The units of the energy are arbitrary since, in this section, we are solely concerned with the relative energy gaps over a normalised annealing time.

Figures \ref{fig:crossing_exampes}(a-c) correspond to three different tri-partite example graph structures. The weights allocated to the three sub-graphs are $W_0=1.05$, $W_1=1.00$ and $W_2=0.95$ and the three sub-graph sizes are given above each plot. The edge penalty is $J_{zz}=5$ (which is the value used throughout the rest of this section.) The plots show the annealing gap spectrum for each instance using the catalyst-free annealing Hamiltonian given by equation \ref{eq:total_H}. A cartoon is included in each case showing the expected driver perturbations, with the perturbed states colour-coded according to the sub-graphs they correspond to. 

Figure \ref{fig:crossing_exampes}(a) shows a case where $G_0$ is larger than both $G_1$ and $G_2$ such that we expect the state corresponding to the global optimum, $\ket{E_0}$, to receive the largest perturbation from the driver. As demonstrated in the cartoon, we therefore expect no PCs to form with the first excited state (1ES). Looking at the associated annealing spectrum, we observe that the smallest gap minimum between the ground and first excited state is approximately equal to that of the problem Hamiltonian, $E_1 - E_0$, indicating the lack of a PC.  

In the second example, given in figure \ref{fig:crossing_exampes}(b), $G_1$ is larger than $G_0$ while $G_2$ is still smaller. Correspondingly, we expect the problem 1ES to receive a larger perturbation from $\hat{H}_d$ than the problem GS such that a PC forms between the instantaneous ground and first excited states. Looking at the annealing spectrum, we can see the associated gap minimum at $s\approx0.9$. We do not expect the perturbed problem 2ES to partake in any PCs and indeed we see that the instantaneous 2ES maintains its separation from the 1ES. 

In the final example, given in figure \ref{fig:crossing_exampes}(c), $G_2$ is larger than both $G_0$ and $G_1$. As such we expect the perturbed problem 2ES to cross both the perturbed 1ES and GS. The two PCs can clearly be seen in the annealing spectrum -- one between the instantaneous 1ES and 2ES and another between the GS and 1ES. 

Finally, let us confirm that the small gap minima appearing in the spectrum do indeed close exponentially as we increase the system size. In figure \ref{fig:crossing_exampes}(d) we plot the scaling of the spectral gap minimum, $\Delta E^\text{min}_{01} = \min_{0 \leq s \leq 1}\left[E_1(s) - E_0(s) \right]$, with system size for two scalable bi-partite graphs. The sub-graph weights in both cases are set to $W_0=1.05$ and $W_1=1.00$. The darker purple data shows the results when $n_0=(n-1)/2$ and $n_1=(n+1)/2$ such that we expect a PC to form. We see that the gap minimum closes exponentially with $n$ as expected. The lighter purple data corresponds to the setting with the sub-graph sizes flipped such that $n_0 > n_1$. In this case we expect no PC to form and indeed we see that the gap closing is sub-exponential. In fact, we find that the size of the gap minimum is approximately equal to $E_1-E_0$ for all $n$. (Data not shown.)

Before moving on, we briefly note that this family of graphs certainly does not encompass all possible instances that would result in the formation of a PC, nor are they intended to. Rather we can think of these instances as concentrating the part of the problem responsible for this annealing bottleneck into a setting that is small enough for practical simulation. 

\subsection{ Single perturbative crossing \label{sec:numerics-1PC} }

\begin{figure*}[]
    \centering
    \includegraphics[scale=0.42]{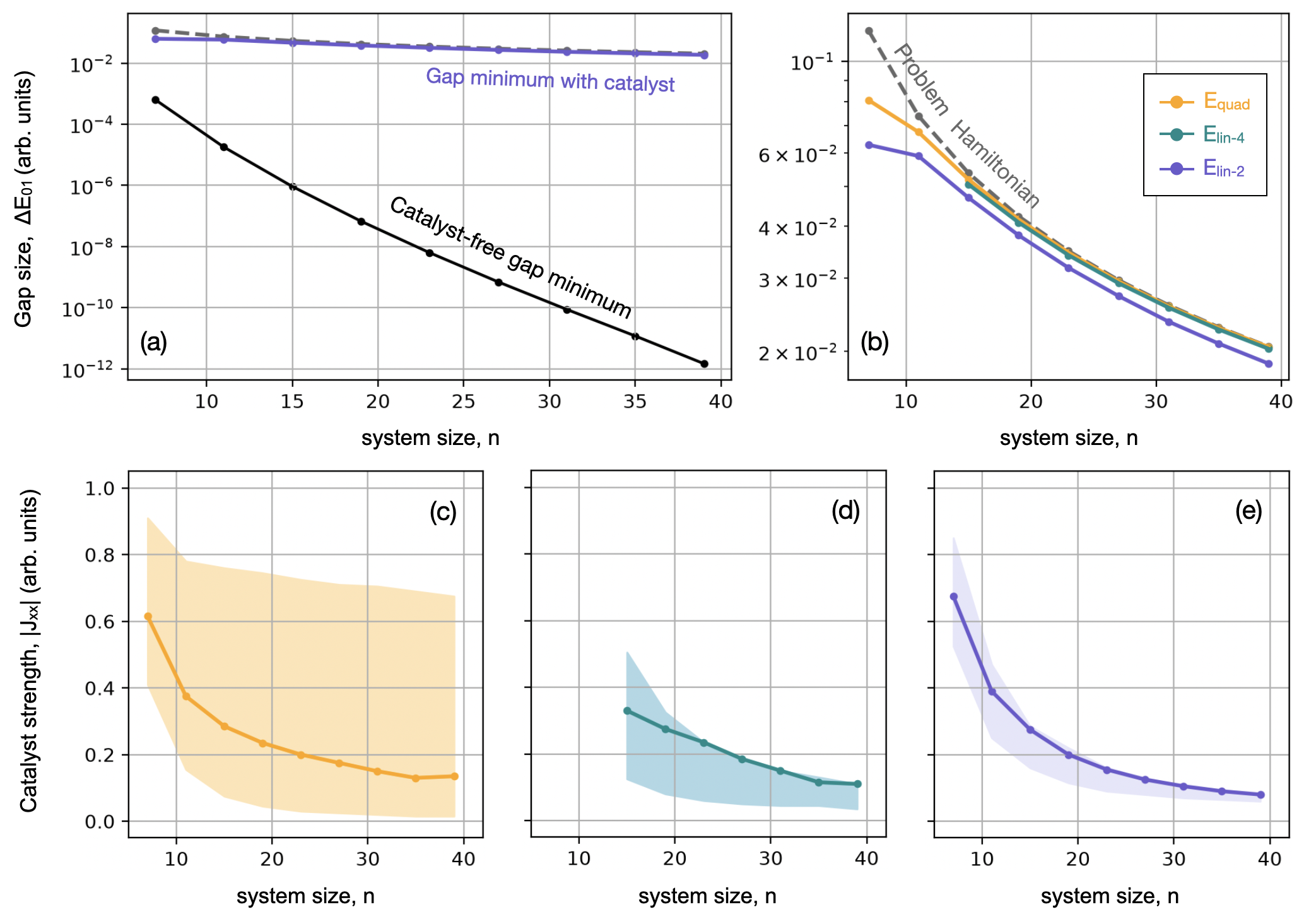}
    \caption{ Plots concerning the performance of the three catalysts outlined in section \ref{sec:numerics-1PC} on a bipartite problem instance with $n_0=(n-1)/2$, $n_1=(n+1)/2$, $W_0=1.05$, $W_1=1.00$ and $J_{zz}=5$ . Figures (a) and (b) show gap scaling data, with (b) showing a close-up view of the catalyst performance data. The solid black curve shows the gap minimum between the instantaneous ground state and first excited state when a catalyst-free annealing schedule is used, the dashed grey curve shows the ground and first excited state separation in $\hat{H}_p$ and the yellow, teal and purple curves show the size of the gap minimum when a catalyst with optimised $J_{xx}$ is introduced with $E_\text{cat}=E_\text{quad}$, $E_\text{lin-4}$ and $E_\text{lin-2}$ respectively. Figures (c-e) give the precision needed in $J_{xx}$ to achieve the demonstrated gap enhancement for $E_\text{cat}=E_\text{quad}$, $E_\text{lin-4}$ and $E_\text{lin-2}$ respectively. The curves show the scaling of the optimal $J_{xx}$ values and the shaded regions show the range of values that achieve $\Delta E^\text{min}_{01} > 0.90 \times \Delta E^\text{opt}_{01}$ where $E^\text{opt}_{01}$ is the optimal gap enhancement and $\Delta E^\text{min}_{01}$ is the achieved enhancement. }
\label{fig:singlePC}
\end{figure*}

In this section we examine the capacity of the proposed catalyst to remove a \textit{single} PC. To this effect, we return to the scalable bipartite graph from the previous sub-section with $W_0=1.05$, $W_1=1.00$, $n_0=(n-1)/2$ and $n_1=(n+1)/2$.

We concluded, in section \ref{sec:XX_selection}, that to remove a PC the $XX-$terms should be selected to couple the excited state responsible for the PC to other low energy states. In the case of the MWIS problem, this can be achieved for one of the local optima by introducing $XX-$couplings between vertices contained in the set. This is because such an $XX-$coupling will introduce a correlated spin flip, effectively removing both vertices from the set and resulting in another independent set. For our family of problem graphs, this amounts to introducing $XX-$terms between vertices in the appropriate sub-graph, $G_a$. For the bipartite problem considered here, we examine catalysts of the form,
\begin{equation}
    \hat{H}_c = J_{xx} \sum \limits_{(i,j) \in E_\text{cat}} \sigma^x_i \sigma^x_j,
    \label{eq:Hc}
\end{equation}
where $E_\text{cat}$ consists of edges between vertices in $G_1$. This catalyst is introduced to the anneal according to equation \ref{eq:total_H_cat}. 

When selecting $E_\text{cat}$, there are $n_1(n_1-1)/2$ potential couplings to choose from and significantly more \textit{combinations} of couplings that we could include. In our numerical simulations, we are somewhat limited in the choices we can make by the fact that disrupting the symmetry of the system reduces the extent to which we can cut down the relevant Hilbert space. We consider three choices for $E_\text{cat}$:
\begin{enumerate}
    \item $E_\text{quad}$: Includes all possible $XX-$terms within $G_1$. In this case the number of couplings included in the catalyst scales quadratically as $n_1(n_1-1)/2 = (n+1)(n-1)/8$.
    \item $E_\text{lin-4}$: This set is produced by separating off four vertices from $G_1$ and coupling those four vertices to every other vertex in $G_1$. This results in a linearly scaling number of $XX-$terms. Specifically, the number of couplings included scales as $4(n_1-4)=2n-15$.
    \item $E_\text{lin-2}$: Same as above except with only two vertices separated off from $G_1$. This results in a linear scaling of $2(n_1-2)=n-3$.
\end{enumerate}
We have listed the options for $E_\text{cat}$ in descending order of how many couplings they introduce. (The reason behind the form of the two linearly scaling choices relates to preserving symmetries within the systems that allow us to reduce the Hilbert space.) For each system size ($7 \leq n \leq 39$) and choice of $E_\text{cat}$ we numerically sweep $J_{xx}$ to obtain the catalyst strength that maximises the gap minimum in the annealing spectrum. 

Figures \ref{fig:singlePC}(a-b) compare the gap scaling with and without the presence of a catalyst, with (b) showing a closer view of catalyst scaling data. The black curve shows the gap scaling in the catalyst-free case. (This is the same data that was presented in figure \ref{fig:crossing_exampes}(d).) The dashed grey curve shows the ground state and first excited state separation in $H_p$. This is the maximum gap enhancement that can be achieved. Finally, the yellow, teal and purple curves show the size of the gap minimum when the catalyst is introduced with $E_\text{cat}=E_\text{quad}$, $E_\text{lin-4}$ and $E_\text{lin-2}$ respectively. We see that, for all three choices of $E_\text{cat}$, the minimum gap in the annealing spectrum is approximately equal to the gap associated with the problem spectrum -- indicating that the PC has ben removed. (Note: the smallest system size differs for $E_\text{lin-4}$ since this catalyst can only be constructed when $n_1>4$ and does not differ from $E_\text{lin-2}$ when $n_1=6$.) 

In figures \ref{fig:singlePC}(c-e) we plot the $J_{xx}$ values that achieve this gap enhancement for the three different choices of $E_\text{cat}$. The solid curves show the $J_{xx}$ values that result in the optimal gap enhancement while the shaded regions show the range of values that result in the gap minimum being at least $0.9$ times the maximised value. The first thing to note is that very small values of $J_{xx}$ are found to produce this $0.9$ enhancement -- on the order of $10^{-2}$. As one might expect, we find that adjusting the parameters associated with the problem (\textit{i.e.}, the weights, $W_a$, and set sizes, $n_a$) changes these magnitudes. A full exploration of how the necessary $J_{xx}$ values depend on the features of the problem that produce the PC we leave for future work. 

With regard to the precision required in $J_{xx}$, we see more prominent differences between the three catalysts than we did with their capacity for gap enhancement, with more couplings in $E_\text{cat}$ reducing the demand on the precision with which $J_{xx}$ must be set. Specifically, the range of $J_{xx}$ values achieving $0.9$ times the optimal gap enhancement is largest for $E_\text{cat}=E_\text{quad}$ and smallest for $E_\text{cat}=E_\text{lin-2}$ across all system sizes. Furthermore, we observe a notable difference in scaling behaviour between the quadratic and the two linearly scaling cases. For $E_\text{quad}$, the $J_{xx}$ range for which the catalyst achieves $0.9$ times the optimal gap enhancement initially increases with $n$ before more or less plateauing while, for both linearly scaling choices, this range decreases.

By extending the analysis from section \ref{sec:theory-removal}, we can shed some light on why the gap enhancement achieved by catalysts containing more couplings might be less sensitive to the precise catalyst strength used. We reserve the full discussion of this for appendix \ref{app:signs}. The key points can however be explained as follows: In section \ref{sec:theory-state_makeup}, we demonstrated that the dominant contributions to the driver perturbed problem state vectors associated with the local optima were positive. This allowed us to deduce that the perturbation from the \textit{catalyst} to these states would have the same sign as $J_{xx}$. These arguments may however break down for sufficiently large $J_{xx}$ since the problem states are now being perturbed by both the driver and the catalyst. The introduction of negative vector components creates a situation where increasing $J_{xx}$ is no longer guaranteed to have a consistent effect on spectrum. In appendix \ref{app:signs} we show that the relevant instantaneous states are more likely to be stable against the appearance of negative vector components if more couplings are included in the catalyst. This can be thought of as competing $XX-$terms introducing frustration with regard to which negative vector components are energetically favourable. As such, we can expect that increasing $J_{xx}$ past $J_{xx}^\text{opt}$ to be less detrimental when more couplings are present in the catalyst.

These same arguments may explain why the performance of the catalyst is weakest on the very smallest system sizes. It is possible that, for these systems, the number of couplings contained in $E_\text{cat}$ is not sufficient for the PC to be fully removed for values of $J_{xx}$ that are small enough to limit the appearance of negative vector components. 

Furthermore, the arguments presented in appendix \ref{app:signs} suggest that the precise interplay between $XX-$terms included in the catalyst may play an important role in the appearance of negative vector components. It may be of potential interest in future work to understand if and how different structures of $E_\text{cat}$ (\textit{e.g.}, a spider graph, pairing up the vertices in $G_1$, or a chain of $XX-$couplings) can improve the robustness of the gap enhancement to changes in $J_{xx}$. 

\subsection{ Multiple perturbative crossings \label{sec:numerics-multiple} }

\begin{figure*}[]
    \centering
    \includegraphics[scale=0.4]{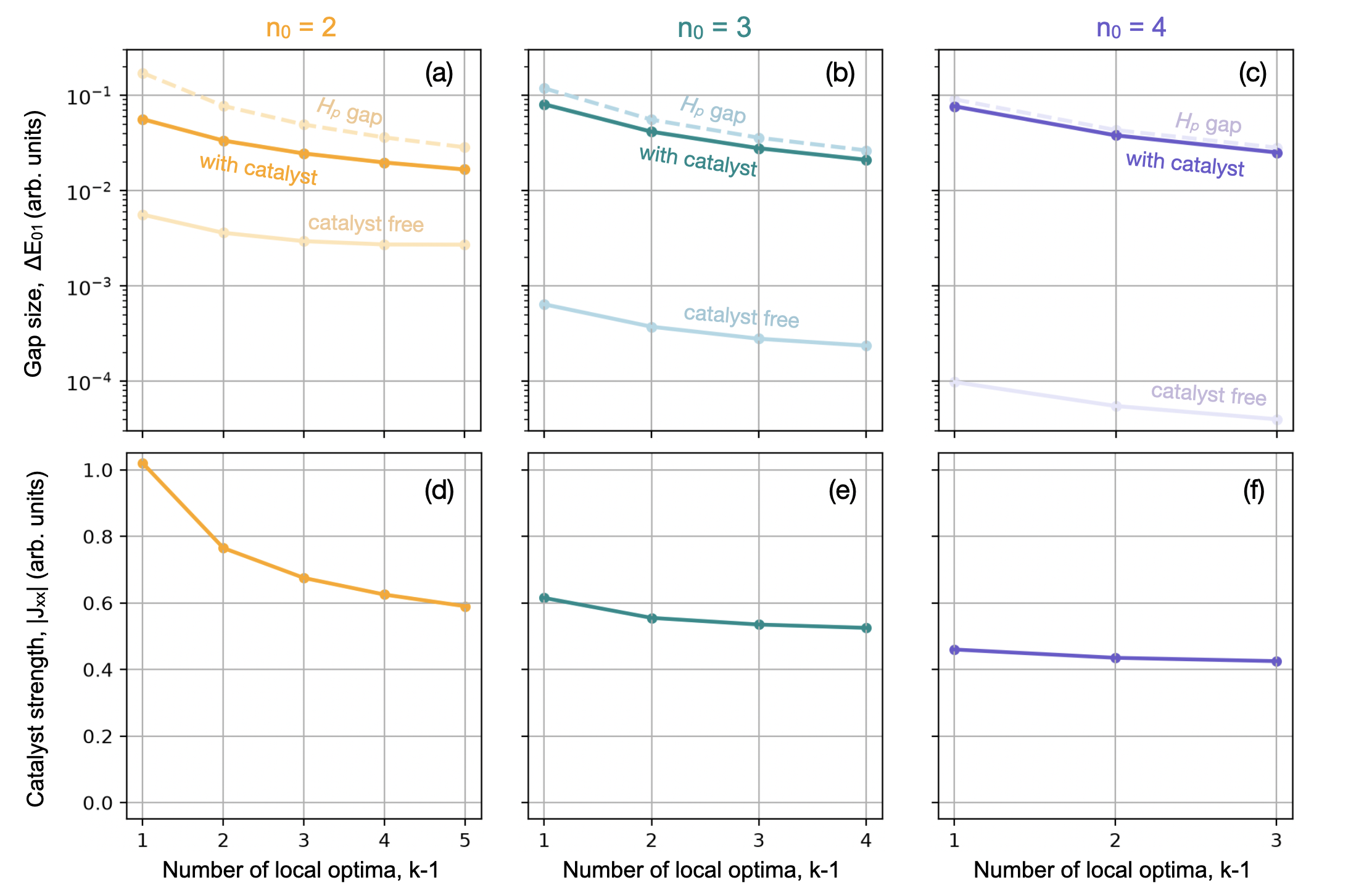}
    \caption{ Plots concerning the performance of the catalyst defined by equation \ref{eq:Hc_mult} on $k$-partite problem instances with $n_a=n_0+1$ $\forall$ $a>0$, $W_0=1.05$, $W_a = 1.00 - 0.01\times(a-1)$ $\forall$ $a>0$ and $J_{zz}=5$. The first, second and third columns of data (presented in orange, teal and purple) correspond to settings with $n_0=2$, $3$ and $4$ respectively. Plots (a-c) show how the size of the gap minimum changes with increasing numbers of local optima ($k-1$). The faint solid curves on each plot show the gap minimum between the instantaneous ground state and first excited state when a catalyst-free annealing schedule is used. The faint dashed curves show the ground state and first excited state separation in $\hat{H}_p$. The opaque solid curves show the size of the gap minimum when a catalyst with optimised $J_{xx}$ is introduced. The optimised $J_{xx}$ values are plotted in figures (d-f). }
\label{fig:multiplePCs}
\end{figure*}

We now demonstrate that the proposed $XX-$catalyst can be used to remove multiple PCs present in the same spectrum, with a separate component of the catalyst targeting each relevant local optimum. 

The graphs we consider in this section consist of $2 \leq k \leq 6$ sub-graphs such that there are $1 \leq k-1 \leq 5$ local optima in addition to the global optimum. Each sub-graph $G_a$ ($a>0$), corresponding to a local optimum, contains $n_0+1$ vertices such that we expect PCs to form between the GS and each of the $k-1$ excited states associated with the local optima. We consider three cases: $n_0=2$, $3$ and $4$. The total weights on $G_0$ and $G_1$ are taken to be $W_0=1.05$ and $W_1=1.00$ as in the bipartite case from the previous section. The weights of the additional sub-graphs are separated by $\Delta W=0.01$ such that $W_a = 1.00 - 0.01(a-1)$ $\forall$ $a>0$. 

Allocating the same sub-graph size, $n_a$, and comparable sub-graph weights, $W_a$, to the sub-graphs, $G_a$ ($a>0$), results in a setting where the properties relevant to the formation of PCs are consistent across the local optima. We find that this allows us to set a universal catalyst strength, $J_{xx}$, rather than allocating a separate catalyst strength, $J_{xx}^a$, to the couplings within each sub-graph. As a result, we only need to optimise a single $J_{xx}$ value for each instance, drastically reducing the demand on the computational resources required to obtain our results. 

We choose to focus our attention, in this section, on catalysts containing all $XX-$couplings within the relevant sub-graphs (\textit{i.e.} $E_\text{cat} = E_\text{quad}$). This choice of $E_\text{cat}$ does not disrupt the symmetry of the original problem and so facilitates the inclusion of more local optima before we reach the limits of what we can simulate.

Thus the form of the catalyst we examine here is
\begin{equation}
    \hat{H}_c = J_{xx} \sum \limits_{a=1}^{k-1} \sum \limits_{(i,j) \in E^a_\text{cat}} \sigma^x_i \sigma^x_j,
    \label{eq:Hc_mult}
\end{equation}
where $E^a_\text{cat}$ contains all $n_a(n_a-1)/2$ possible $XX-$couplings between vertices in $G_a$. The index, $a$, in the first sum runs over all sub-graph labels (other than $a=0$ which is associated with the global optimum) such that couplings are introduced to target each of the $k-1$ local optima. 

Figures \ref{fig:multiplePCs}(a), (b) and (c) show the effect on the gap minimum when this catalyst is introduced into the setting we have described with $n_0=2$, $3$ and $4$ respectively. Each plot shows the change in the spectral gap minimum, with and without the catalyst, as additional local optima are added. As in the previous section, $J_{xx}$ is swept individually for each system and the optimal $J_{xx}$ values are plotted below the corresponding gap data in figures \ref{fig:multiplePCs}(d-f). Since the size of the Hilbert space still scales exponentially with the number of sub-graphs, $k$, we are limited in the number of local optima we can introduce. For the $n_0=2$, $3$ and $4$ cases we are able to introduce up to $5$, $4$ and $3$ local optima respectively. 

\begin{figure*}[]
    \centering
    \includegraphics[scale=0.33]{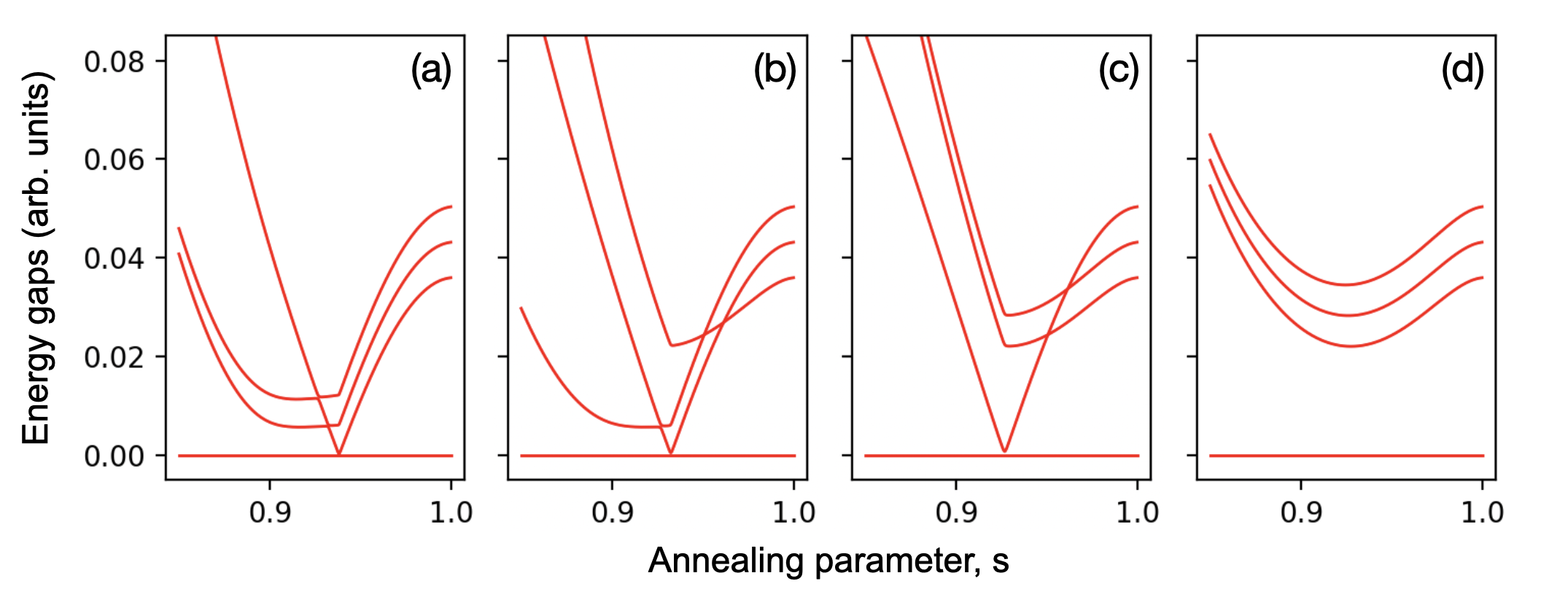}
    \caption{ Gap spectra associated with the problem instance defined by $n_0=3$, $n_1=n_2=n_3=4$, $W_0=1.05$, $W_1=1.00$, $W_2=0.99$, $W_3=0.98$ and $J_{zz}=5$. Plot (a) shows the catalyst-free spectrum and plots (b-d) show the spectrum in the presence of three different catalysts containing all-to-all $XX-$couplings within the following sub-graphs: (b) $G_1$ (c) $G_1$ and $G_2$ (d) $G_1$, $G_2$ and $G_3$. The catalyst strength is set to $J_{xx}=0.535$ which was the value found to maximise the gap minimum. }
\label{fig:removal_example}
\end{figure*}

Note that the catalyst-free gap minimum does not appear to close exponentially with the number of local optima. This is to be expected since the exponential gap closing generally associated with an avoided level-crossing comes from an increase in Hamming distance between the relevant pair of states. This can generally be expected to grow with the system size. Here, however, the Hamming distance between the global optimum and any one of the local optima is consistently $2n_0+1$. Thus, what we are looking for here is not a suppression to an exponential gap closing, but rather a consistent capacity to enhance the gap minimum with increasing numbers of PCs present in the original annealing spectrum.

Looking at the results in each plot we observe that the catalyst is indeed able to enhance the gap minimum in all cases and that its effectiveness is not hampered by the inclusion of additional PCs that the catalyst must target. We also observe that as we scale the system, \textit{i.e.}, move from plot (a) to (c), that the gap enhancement becomes more pronounced. This is in line with our findings from the previous section and can again be accounted for by an increase in the number of $XX-$couplings that target each of the local optima. 

Finally, to demonstrate the importance of each set of couplings, and also to clearly illustrate the removal of several PCs, we present a set of spectra in figure \ref{fig:removal_example} showing the successive removal of three PCs. These spectra correspond to the problem instance with $n_0=3$ and $n_1=n_2=n_3=4$; \textit{i.e.}, the third data point in figure \ref{fig:multiplePCs}(b). Figure \ref{fig:removal_example}(a) shows the catalyst-free spectrum towards the end of the anneal. Figure \ref{fig:removal_example}(b) then shows the results when $XX-$couplings within $G_1$ are included such that the PC with the first excited state is removed. Figures \ref{fig:removal_example}(c) and (d) show the results when the couplings within $G_2$ and $G_3$ are added into the catalyst as well, such that the remaining PCs are also removed. 

\section{ Recursive protocol for $XX-$catalyst construction \label{sec:algorithm} }

This section introduces and examines a recursive approach to catalyst-construction based on the findings of the preceding sections. In the toy problems examined so far, we have known in advance where the $XX-$couplings should be placed in order to remove the PC(s) for a given instance. However, we have also suggested that it may be possible to obtain this information for more general problems from ``failed'' annealing runs. This is because the annealer is likely to return precisely the local optima responsible for the PCs present in the spectrum. 

We begin, in section \ref{sec:algorithm_description}, by proposing a recursive approach where a catalyst is iteratively updated based on the measurements returned by successive annealing runs, first discussing the protocol in its most general form before going on to describe the process that we will be simulating. The instance set on which we test this approach is outlined in section \ref{sec:algorithm_instances}. We present and discuss our numerical results in section \ref{sec:algorithm_results}. Overall, we find that this approach to catalyst-construction allows the problem GS to be found significantly faster than if no catalyst or randomly chosen $XX-$terms were used. 

\subsection{ Protocol Description \label{sec:algorithm_description} }

The aim of the protocol is to construct a catalyst with the form given by equation \ref{eq:Hc}, defined by the edge set, $E_\text{cat}$ and coupling strengths, $J^{xx}_{i,j}$. The protocol begins with an empty set of edges and proceeds in iterations with each iteration consisting of four steps:
\begin{enumerate}
    \item $N_\text{runs}$ calls are made to the quantum annealer, each with an annealing time of $T$, to obtain $N_\text{runs}$ candidate states.  
    \item The lowest energy state is selected. If this is not a local optimum, gradient descent is used to obtain a local optimum. 
    \item A termination condition is assessed to determine whether or not to perform further iterations. If the termination condition is met, the best solution obtained over the entire protocol is returned.
    \item If the termination condition is \textit{not} met, the catalyst is updated based on the local optimum obtained by the current iteration.
\end{enumerate}
The crux of the proposed protocol is that we are assuming that the obtained optima are, if not the global optimum, the local optima in which QA is becoming trapped as a result of PCs. Step 4 should, therefore, update the catalyst in such a way that we introduce $XX-$terms that couple the local optimum identified at that iteration to other low energy states. 

While calculating the full eigenspectrum of $\hat{H}_p$ is exponentially computationally expensive in the problem size, the complexity of calculating the energy of a single known problem eigenstate scales only linearly. There are $n(n-1)/2$ potential $XX-$couplings to choose from. Thus, given a state returned by the annealer, the time it would take to check the energy of every state it could be $XX-$coupled to scales as $n^3$. As such, it is feasible to check all possible interactions and select the ones that couple the identified local optimum to the lowest energy state. 

This approach to selecting candidate catalyst terms has the benefit of being entirely general. However, the computational demand of updating the catalyst could be reduced further by using the nature of the problem type to guide the placement of $XX-$interactions. For instance, we noted that a local optimum in the MWIS problem can be easily coupled to low energy states by introducing $XX-$interactions between vertices that are in the set. This observation negates the need to actively identify the states being coupled to and calculate their energies. It is possible that similar strategies could be identified for other problem types. 

Let us now consider step 3, in which a termination condition is assessed. In practice, it is not possible to know if and when the global optimum to a problem has been found. Any termination condition must therefore be based on other, accessible, factors. Examples of what these could include are a particular quality of solution being reached, a maximum number of iterations, or the same state being measured a certain number of times. 

We now introduce the specifics of the protocol that we simulate in section \ref{sec:algorithm_results}, beginning with the process for updating the catalyst in step 4. For the purpose of seeing how a general approach performs, we will be selecting the couplings based on the energies of the associated states rather than utilising the structure of the MWIS problem. The specific steps, given an obtained state $\ket{E_a}$ corresponding to a local optimum, are as follows:
\begin{itemize}
    \item Generate a list of the $n(n-1)/2$ potential states that $\ket{E_a}$ could be coupled to by the $n(n-1)/2$ candidate $XX-$terms. Use $\hat{H}_p$ to calculate the energies of these states.
    \item Select the $XX-$terms associated with the $M \times N_{XX}$ lowest energy states ($M \in \mathbb{Z}^+$). Then, from this list, select $N_{XX}$ couplings at random.
    \item For each of these couplings, $(i,j)$, if the $XX-$term is already present in the $E_\text{cat}$, the associated strength is increased by $\Delta J_{xx}$. If the coupling is not already present in $E_\text{cat}$, it is introduced with a strength of $\Delta J_{xx}$.
\end{itemize}
This process allows for measurements to both guide the selection of coupling locations and tune the strengths of these couplings. Smaller values of $\Delta J_{xx}$ allow for finer grained tuning at the cost of a greater numbers of iterations being needed to reach catalyst strengths large enough for a PC to be lifted. On the other hand, larger values of $\Delta J_{xx}$ might result in a suitable coupling strength being overshot. The parameter $N_{XX}$ sets the number of $XX-$terms that are selected at each step. If $M=1$, the $XX-$terms associated with the $N_{XX}$ lowest energy states will be chosen. When $M>1$, we select from among the $M \times N_{xx}$ lowest energy states. The motivation for the parameter $M$ is to introduce some variation into the selection process even when the same local optimum is obtained more than once. We elaborate on why this might be important in section \ref{sec:algorithm_instances}.

Since, in our simulations we have access to the full problem spectrum, we are free to use the sampling of the GS as part of the termination condition. While this is not realistic, it facilitates our goals here which are to understand how effectively a recursively updated catalyst can guide QA to the global optimum. We also introduce a second termination condition which is that the number of iterations does not exceed $I_\text{max}$. 

A specific realisation of the proposed protocol is defined by six parameters: $N_\text{runs}$, $T$, $N_{XX}$, $M$, $\Delta J_{xx}$ and $I_\text{max}$. We will introduce the specific parameter choices we make use of in our investigation at the start of section \ref{sec:algorithm_results}.

\subsection{ Quasi-random MWIS instances \label{sec:algorithm_instances} }

\begin{figure*}[]
    \centering
    \includegraphics[scale=0.4]{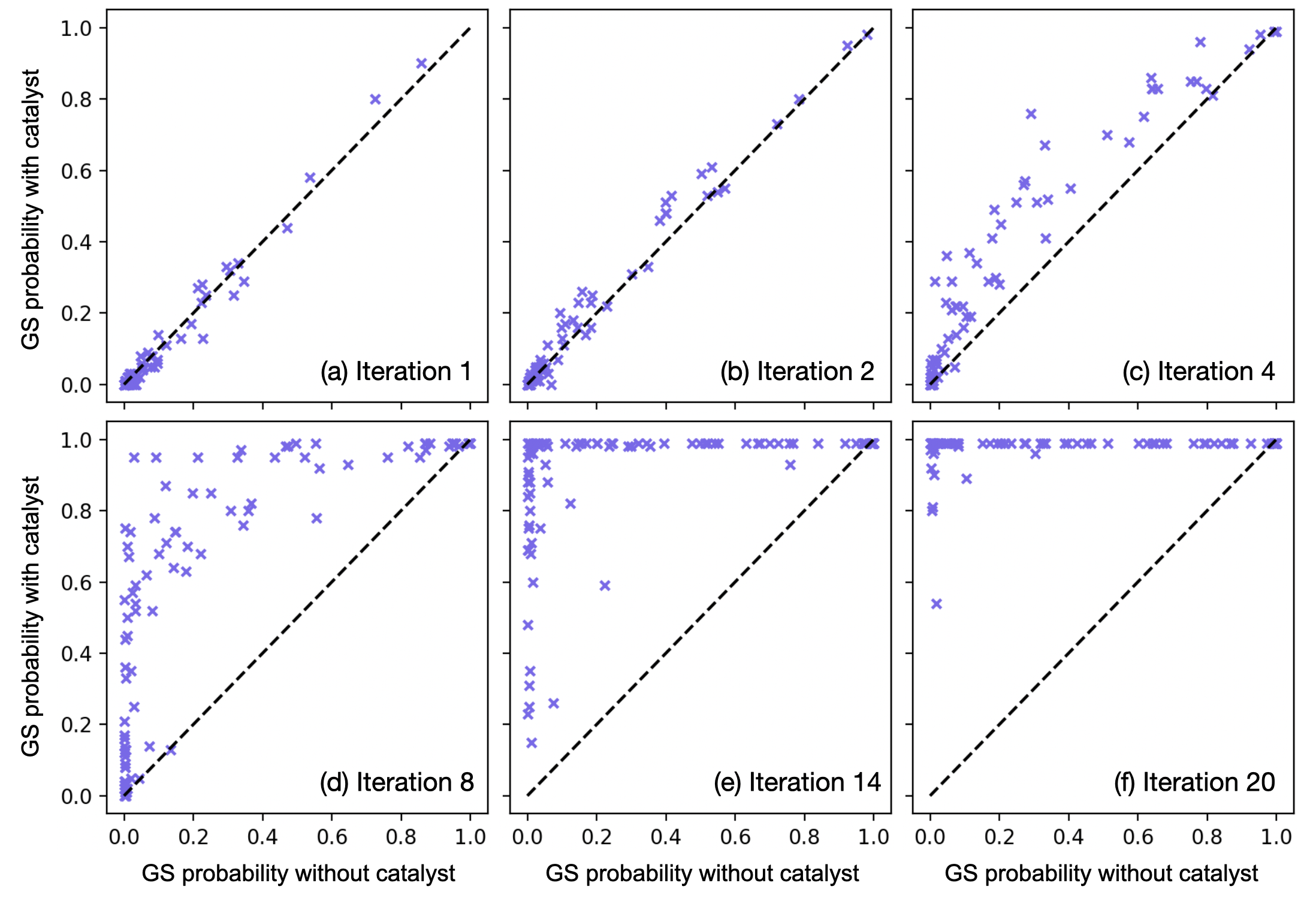}
    \caption{ Plots comparing the probability of GS measurement with and without a catalyst with each plot corresponding to a different iteration. For a given plot, each cross corresponds to a particular instance. The GS probability without a catalyst is calculated directly from the final GS overlap of the catalyst-free anneal. The GS probability \textit{with} the catalyst is calculated as the fraction of runs that had found the GS by that iteration. }
\label{fig:alg_steps}
\end{figure*}

The highly structured instances described and utilised in section \ref{sec:numerics} were introduced to facilitate a straightforward investigation of the scaling behaviour of the proposed catalyst. In addition to allowing us to easily produce clearly defined PCs, the symmetries present in these systems allowed us access to larger system sizes. We now want to move to more general MWIS settings to understand if a targeted, recursively constructed, $XX-$catalyst can remove PCs from such settings.

The challenge in creating a suitable set of problem instances is in finding a balance between generality, simulability and hardness for QA. Losing the symmetries present in the structured instances means we are restricted to smaller system sizes ($18$-spin graphs). There are various potential options for random graph generation such as Erdős–Rényi and Barabási–Albert. However, at the system sizes we consider, these methods do not readily produce MWIS problems whose QA spectra contain prohibitively small gaps -- let alone QA spectra with \textit{multiple} PC features. Thus, such approaches to problem generation will not provide us with the kind of settings we are interested in. Carefully chosen structures and vertex weights can help us produce instances that are hard for QA, even for small systems. However, introducing too much structure limits the expressivity of our instance set.

To strike a balance between these factors, we have chosen to do the following. Rather than randomly generate the \textit{graph structure}, we randomly generate the properties of $3 \leq N_l \leq 7$ local optima, in addition to a global optimum. The total weight and number of vertices associated with each of these local optima are randomly selected. The global optimum and $N_l$ local optima are then randomly seeded within an $18$-spin graph by setting the vertex weights and edges of the graph chosen accordingly. A full description of how the properties of the optima are chosen and how the graph is constructed around them is given in appendix \ref{app:instance_construction}. 

Other than the greater variety in weights and graph structure that this gives us, a crucial difference between the instances generated through this method and those utilised in section \ref{sec:numerics} is that vertices can be shared between multiple local optima. Aside from giving these instances greater expressivity, this brings us to the reason for the parameter, $M$, from the preceding section which introduces some randomness to the catalyst terms introduced based on a given local optimum. 

Recall that we want to select $XX-$terms that couple the identified local optimum to other low energy states. This will raise the energy of the associated perturbed state, bringing it away from the perturbed problem GS. Consider a setting where the local optimum we have identified shares $n_s$ vertices with the global optimum, whose structure we assume we do not know, and that these $n_s$ vertices are lower in weight than the other vertices contained in the local optimum. If we select $XX-$terms which couple the local optimum to the \textit{lowest possible} energy states, terms coupling these $n_s$ vertices will be selected first (since this amounts to removing the lowest-weighted vertices from the set). 

While these couplings choices will indeed increase the energy of the perturbed excited state, they will \textit{also} increase the energy of the perturbed problem GS -- therefore limiting the effectiveness of these coupling choices. However, there will always be vertices that are not shared between a given local and the global optimum. By selecting from \textit{amongst} the lowest energy choices rather than the \textit{very} lowest energy choices, we allow for the possibility of finding these un-shared couplings. 

We generated $89$ problem instances which we utilise in the following sub-section. We make no comment on the computational hardness of this problem set. Its purpose is simply to provide a variety of conditions under which multiple PCs can form. 

\subsection{ Results \label{sec:algorithm_results} }

\begin{figure*}[]
    \centering
    \includegraphics[scale=0.34]{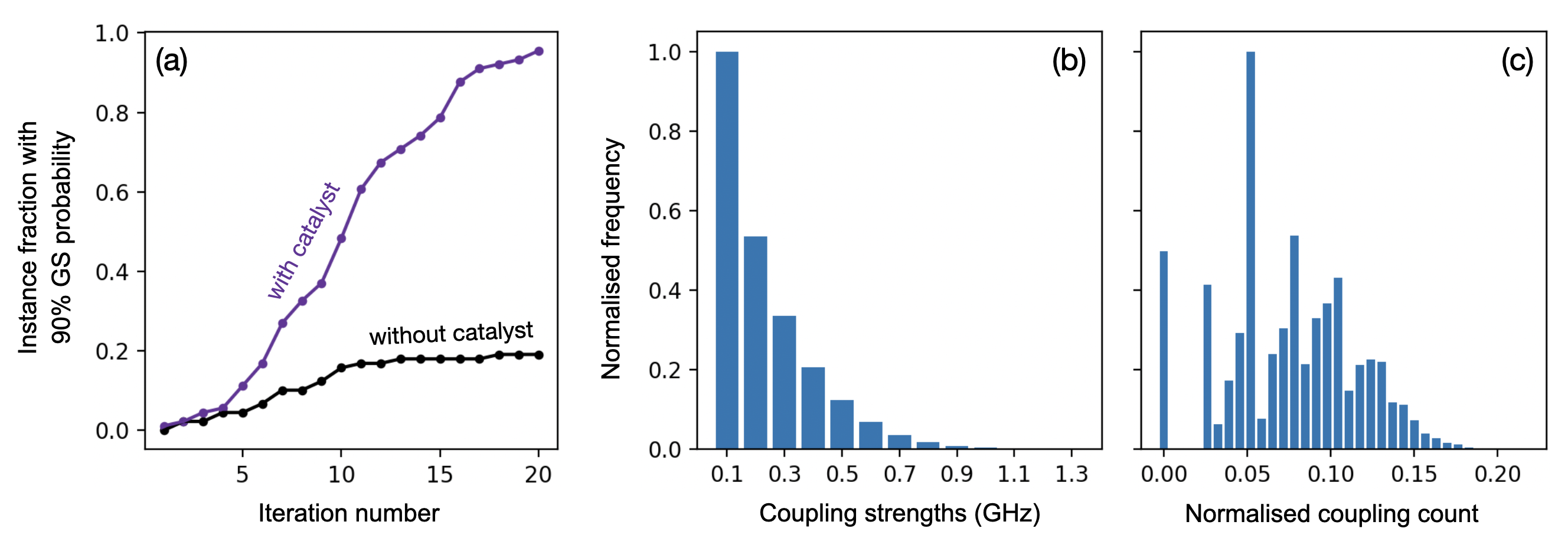}
    \caption{ Plots providing summary statistics associated with the proposed protocol discussed in section \ref{sec:algorithm}. (a) shows how the fraction of instances solved with $90\%$ probability evolves with iteration number. The results with and without an updating catalyst are shown in purple and black respectively. Plots (b-c) illustrate the properties of the constructed catalysts which resulted in GS measurements. (b) shows the strengths with which the couplings are introduced and (c) shows the number of couplings present in the catalysts as a fraction of the total number of potential coupling locations, $153$. }
\label{fig:alg_summary}
\end{figure*}

To simulate the process outlined in section \ref{sec:algorithm_description} for a given instance, we replace the $N_\text{runs}$ calls to a quantum annealer with a closed system simulation of the quantum annealing process, sampling the final state $N_\text{runs}$ times to obtain the ``measured'' eigenstates of $\hat{H}_p$. 

To quantify the performance of the protocol on a single problem instance, we simulate the process $100$ times. In these simulations we use an annealing time of $T=60\text{ns}$. With the fields and couplings being introduced on the order of $1\text{GHz}$, this gives a scenario comparable to QA on the D-Wave -- though with the addition of hypothetical $XX-$interactions. For simplicity we take the number of QA calls at each catalyst iteration, $N_\text{runs}$, to be $1$. The number $XX-$terms selected at each iteration, $N_{XX}$, is chosen to be $4$ and we set $M=2$. The catalyst strength increment, $\Delta J_{xx}$, is set to $0.1$. Finally, the maximum iteration number, $I_{max}$, is set to $20$ -- which we find to be sufficient to obtain the GS with high fidelity for this instance set. In future work, it may be of interest to see how tuning the parameters $N_\text{runs}$, $N_{XX}$, $M$ and $\Delta J_{xx}$ affects performance.

For each instance, we give the probability of measuring the GS by iteration $I$ as the fraction of times the protocol had found the GS by this iteration. We denote this by $P_\text{cat}(I)$. For the catalyst free case, the probability of a GS measurement by each iteration is calculated exactly as
\begin{equation}
    P_\text{cat-free}(I) = 1 - (1-|\braket{E_0|\Psi(s=1)}|^2)^I.
\end{equation}
Since the first iteration of the proposed protocol runs a catalyst-free anneal, we expect  $P_\text{cat}(1)$ to be approximately equal to $P_\text{cat-free}(1)$. These values are not exactly equal since $P_\text{cat}(I)$ is obtained through a probabilistic process.

Figures \ref{fig:alg_steps}(a-f) compare $P_\text{cat}(I)$ (y-axis) and $P_\text{cat-free}(I)$ (x-axis) at iterations $I=1$, $2$, $4$, $8$, $14$ and $20$. We see that, as we move to larger iterations, the points migrate towards the top of the plot, indicating that the updating catalyst Hamiltonian is helping guide the system towards the GS. After $20$ iterations of catalyst updates, the probability of measuring the GS was close to unity for the majority of instances and over $50\%$ for every instance tested. Of particular note is the catalyst performance on the very hardest instances -- which are those of the very left hand side of the plots. We see that, even where repeated catalyst-free annealing runs result in a negligible probability of GS measurement, the recursively constructed catalyst is able to bring this probability to above $80\%$ in almost all cases. 

The overall performance difference between QA with and without the updating catalyst method is illustrated in figure \ref{fig:alg_summary}(a). Here we show the fraction of instances with a $90\%$ chance of GS measurement with increasing iteration number. For the catalyst-free case, the fraction slowly increases and plateaus at around $0.2$. When using an updating catalyst however, the increase in instance fraction is significantly sharper and approaches unity by iteration $20$.

So far, we have established that recursively introducing $XX-$terms into a catalyst Hamiltonian helps quantum annealing to find the GS of our test instances faster than simply repeating catalyst-free annealing runs. To confirm that the \textit{placement} of the couplings is integral to this improvement, we compare the performance of what we will call the \textit{measurement-guided} catalysts produced by the recursive protocol to that of random catalysts with comparable numbers of couplings and values of $J^{xx}_{i,j}$. To understand the effect of the catalysts on the annealing spectra, we will now be considering the final GS overlap achieved by a \textit{single} anneal --  rather than overall probability of measuring the GS after multiple steps as we did in figure \ref{fig:alg_steps}.

\begin{figure*}[]
    \centering
    \includegraphics[scale=0.34]{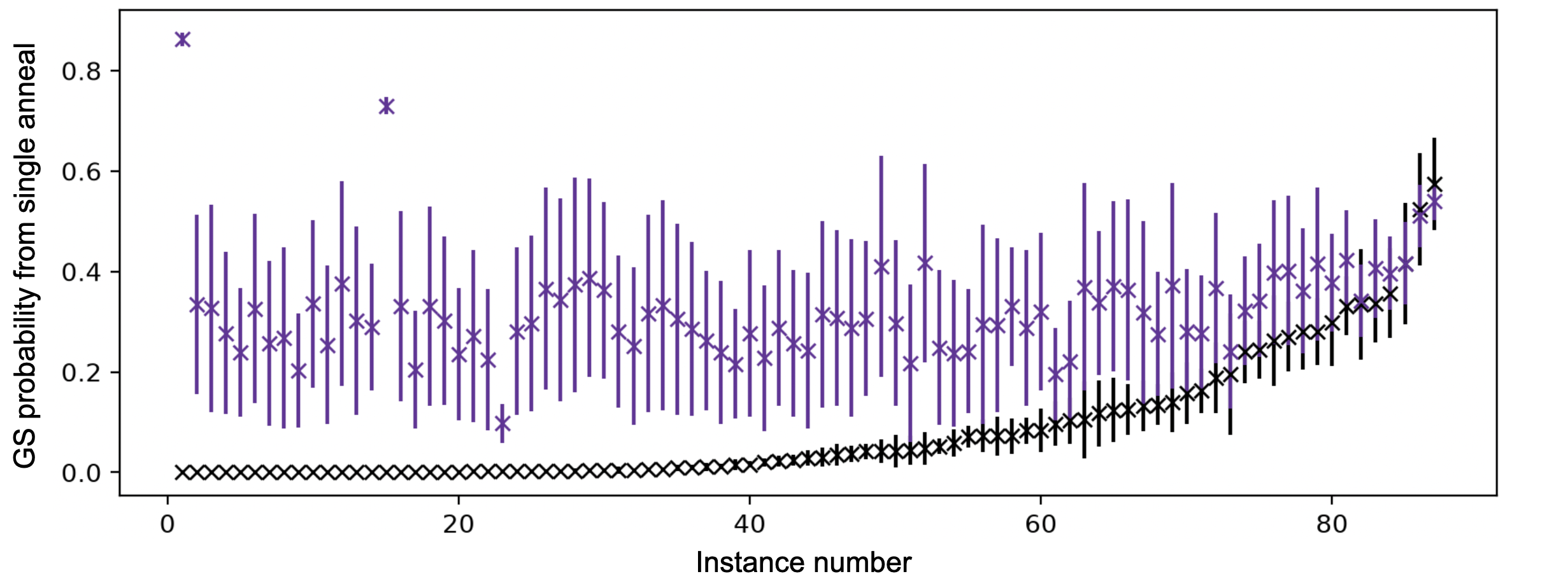}
    \caption{ Plot comparing the final GS overlaps achieved by the measurement-guided catalysts (purple) to comparable randomly generated catalysts (black) across the different problem instances. The crosses show the mean GS overlap obtained after a single anneal and the range shows the standard deviation. The problem instances have been ordered by the mean GS overlap achieved by the randomly generated catalysts. }
\label{fig:rand_cat_comp}
\end{figure*}

To obtain the statistics we use to generate our random catalysts for a given instance, we consider all $100$ measurement-guided catalysts generated by the $100$ runs of the protocol for that instance. Where a run of the protocol led to a GS measurement, we take the catalyst produced at the step the GS measurement occurred. Where the run did not lead to a GS measurement, we use the catalyst as obtained by step $20$. For each randomly generated catalyst, the number of couplings is selected by sampling from a normal distribution defined by the mean and standard deviation of the number of couplings present in the measurement-guided catalysts for that particular instance. The couplings themselves are selected at random from the list of all potential couplings. To set the strength of the randomly generated catalyst couplings, we calculate the mean total catalyst strength for the instance as,
\begin{equation}
    \text{mean}(J^{xx}_\text{tot}) = \sum \limits_{c \in C} \sum \limits_{(i,j) \in E_c} J^{xx}_{i,j}/100
\end{equation}
where $C$ is the set of $100$ measurement-guided catalysts and $E_c$ is the set of edges included in catalyst $c \in C$. For each randomly generated catalyst, we split $J^\text{tot}_{xx}$ for the given instance evenly between the couplings.

For each instance, we generate $100$ random catalysts to compare to the $100$ measurement-guided catalysts generated by the protocol. The mean GS overlap and its standard deviation is plotted for each instance in figure \ref{fig:rand_cat_comp}, with the statistics of the measurement-guided and randomly generated catalysts plotted in purple and black respectively. For visual ease, the instances are ordered by the mean GS overlap obtained by the randomly generated catalysts. Looking at figure \ref{fig:rand_cat_comp}, we see that the measurement-guided catalysts decisively outperform the randomly generated ones for the majority of instances. We also observe that the performance of the measurement-guided catalysts is fairly consistent, with the majority of the mean GS overlaps being between $0.2$ and $0.4$. That this remains the case for the instances where the randomly generated catalysts consistently achieve a negligible GS overlap highlights the effectiveness of utilising available information to make targeted coupling selections over simply introducing more interactions.

So far, we have assumed access to a quantum annealing device that could implement arbitrary $XX-$couplings. In practice, any future quantum annealer that is able to implement $XX-$interactions (either directly or through effective means) is bound to have restrictions to connectivity and interaction strength. To start to understand the requirements on such a device for the purpose of implementing the catalyst approach explored in this section, we present some summary statistics on the catalysts obtained in figures \ref{fig:alg_summary}(b-c).

Figure \ref{fig:alg_summary}(b) shows the normalised frequency of the different coupling strengths, $J^{xx}_{i,j}$, that enter the measurement-guided catalysts. We count each coupling in every catalyst responsible for a successful GS measurement. We see that the very largest catalyst strengths do not go significantly above $1\text{GHz}$ -- \textit{i.e.}, enter on the same order as the homogeneous $X-$field as well as the local $Z-$terms and $ZZ-$couplings. 

Next, figure \ref{fig:alg_summary}(c) shows the normalised frequency of catalyst sizes. The catalyst sizes are given as a fraction of the total number of possible coupling choices which is $153$. We count every catalyst that was responsible for a successful GS measurement across all instances. The very largest catalysts contain no more than a fifth of the total potential couplings with the majority of catalysts containing significantly fewer. For comparison, the $ZZ-$connectivity on the D-Wave advantage 2 is on the order of $1/200^\text{th}$ of a fully-connected graph. It is worth noting, however, that the significantly larger fraction of required $XX-$interactions observed here may be an artifact of the small systems we examine, in which one coupling is already more than $1/200^\text{th}$  of the total potential couplings. Furthermore, as we already discussed with the structured instances of section \ref{sec:numerics}, we are essentially concentrating the hardest parts of the MWIS problem into a setting small enough to simulate. It is possible that the fraction of $XX-$interactions is reflective of that.

\section{ Discussion \label{sec:conclusion} }

In this paper we have proposed a protocol for recursive catalyst-construction that utilises accessible information provided by successive annealing runs. We have performed a numerical proof-of-concept study on a set of quasi-randomly generated MWIS instances that demonstrates the effectiveness of the approach. These results are supplemented by numerical results for a toy model which we used to investigate the scaling behaviour of the catalysts. 

For $95\%$ of the instances examined, the proposed protocol was able to obtain the global optimum of the problem with over $90\%$ fidelity. This is compared to standard annealing for which a $90\%$ fidelity was only achieved for $20\%$ of the instances after the equivalent number of calls to the annealer. We also found that the targeted catalysts generated by this process significantly out-performed the randomly generated ones for all but the very easiest to solve (for QA) instances. 

In addition to validating the proposed method of $XX-$term selection, these findings provide further evidence for how critical the choice of coupling locations is, motivating further work in this direction. Furthermore, we argue that these results provide an optimistic outlook, in that the number of $XX-$couplings required to see the kind of enhancements reported by other works may well be significantly fewer than those used, since a much smaller sub-set of those couplings may be responsible for the majority of the gap enhancement. 

Current QA hardware does not currently natively support $XX-$interactions. However there are a number of proposals for the implementation of such interactions both on superconducting devices \cite{zhang2023tunableinductivecouplerhigh,PhysRevApplied.13.034037} and Rydberg atom systems \cite{Scholl_2022}. In addition to this, there have been a number of proposals for how \textit{effective} $XX-$interactions could be implemented using more readily available hardware \cite{banks2025gadgetssimulatingnonnativexx,Headley_2026,dadgar2026emulatingxxcatalystsquantum,banks2026enhancedanaloguequantumcomputation}. All of these methods come with their own benefits and drawbacks as well as limitations to both the coupling strength and their placement within a larger graph structure, further highlighting the importance of knowing what $XX-$interactions will produce the greatest contributions to gap enhancement. By carefully selecting the $XX-$terms that enter the Hamiltonian, we may be able to significantly reduce the hardware demands on any $XX-$coupling strategy, be that a physical hardware element or a method for implementing effective $XX-$terms.

Non-stoquastic $XX-$interactions, such as those studied in this work, are harder to implement than stoquastic ones. For example, the method proposed in \cite{banks2026enhancedanaloguequantumcomputation} would require non-stoquastic local driving to produce a non-stoquastic effective $XX-$interaction. However, the use of non-stoquastic Hamiltonians as a resource for quantum enhancement is appealing due to this class of Hamiltonians being harder to simulate classically with, for instance, Quantum Monte Carlo methods encountering the sign problem \cite{gupta2019elucidating}. They also introduce the capacity for universal quantum computation in the adiabatic setting \cite{aharonov2005adiabatic}. 

So far, we have only examined the catalyst-construction protocol on a small set of $18-$spin instances. The process for constructing these instances was specifically designed to produce problems whose annealing spectra would contain multiple PCs despite being limited to small system sizes when simulating such non-symmetric instances. While providing a suitable initial test-bed, our simulations of the protocol on this instance set do not allow us to make any claims regarding scaling improvements over catalyst-free annealing. Rather these results provide an initial demonstration that this method for catalyst-construction is indeed successful at removing the PCs present in a spectrum. A natural next step would be to characterise the scaling behaviour of different features of the protocol. Primarily, it will be of interest to understand how the scaling of the number of QA calls needed to reach a given GS fidelity using our catalyst-construction method compares to the number of catalyst-free QA calls required. 

While the computational challenges involved with the simulation of general quantum systems mean that we have not yet examined the scaling behaviour of the protocol, the numerical results on our scalable \textit{toy} model, presented in section \ref{sec:numerics}, offer an encouraging outlook. These toy models are, in essence, a means of producing the kind of annealing bottlenecks (\textit{i.e.} PCs) that are commonly encountered as part of a larger system. Examining the effect of targeted catalysts on a single PC, we found that a linearly scaling number of couplings was sufficient to remove the bottleneck up to system sizes of $39$ spins. We also did not observe any suppression of the catalyst's ability to enhance the minimum spectral gaps as we increased the number of PCs which the catalyst was required to remove. 

As part of our proposal, we introduced a number of parameters associated with the catalyst-construction protocol. These included the number of calls to the annealer per catalyst iteration, $N_\text{runs}$, the number of couplings selected at each iteration, $N_{XX}$, and the catalyst strength increment, $\Delta J_{xx}$. It may be of interest in future work to investigate how the performance of the protocol can be optimised by tuning these parameters. For our toy model we found that less precision in the catalyst strength was needed if more couplings were introduced. This suggests that increasing $N_{XX}$ may allow for the selection of larger $\Delta J_{xx}$ values without risking overshooting suitable couplings strengths, thereby reducing the number of iterations needed for a sufficiently large (but not too large) coupling strength to be reached. Such considerations are likely to be highly setting specific, both in terms of the nature of the optimisation problem and also the hardware limitations.

That the theoretical motivation for the coupling choices stems from features of the problem spectrum, rather than specific graph structures or characteristics of the MWIS problem, gives us reason to believe that this strategy should apply to more general MWIS instances as well as other problem types. While this, alongside our numerical data, offers an optimistic outlook for this approach to catalyst-construction, it remains to be seen precisely how well the protocol performs in more general settings, and to what extent we can expect a scaling advantage over catalyst-free annealing and potentially classical optimisation approaches. 

\begin{acknowledgments}
This work is supported by EPSRC (grant ref. EP/Y004590/1) and the  QCi3 Hub, funded by EPSRC (grant ref. EP/Z53318X/1). The authors also gratefully acknowledge the Gauss Centre for Supercomputing e.V. (www.gauss-centre.eu) for funding this project by providing computing time on the GCS Supercomputer JUWELS~\cite{Juwels_Cluster} at Jülich Supercomputing Centre (JSC).
\end{acknowledgments}

\bibliography{refs}

\appendix

\section{ Signs of the ground state vector components \label{app:signs} }

In this appendix, we examine (a) how the relative signs in the instantaneous GS vector are affected by the introduction of different $XX-$terms and (b) what impact we can expect this to have on their capacity for gap enhancement. This is a truncated version of the discussion found in \cite{feinstein2025}. It makes use of an energetic argument introduced in \cite{Choi2021} where the expectation value of the total Hamiltonian is used to motivate the introduction of specific $XX-$couplings in order to achieve particular relative signs between certain GS vector components. In our case, we want to do the reverse. That is, we consider the introduction of $XX-$terms as motivated in section \ref{sec:theory-removal} and want to understand what this means for the relative signs with which the different problem states contribute to the instantaneous GS. We supplement our theoretical discussion with numerical results for an example graph from section \ref{sec:numerics-setting}. 

The expectation value of $\hat{H}(s)$ for the instantaneous ground state is given by
\begin{multline}
    \braket{\hat{H}(s)} = \bra{E_0(s)}\hat{H}(s)\ket{E_0(s)} = 
    \\-(1-s)\sum \limits_{(a,b) \in C_{x}} \braket{E_0(s)|E_a} \braket{E_b|E_0(s)} 
    \\ + s (1-s) J_{xx} \sum \limits_{(a,b) \in C_{xx}} \braket{E_0(s)|E_a} \braket{E_b|E_0(s)} 
    \\ + s \sum \limits_{a=0}^{2^n-1} E_a |\braket{E_0(s)|E_a}|^2,
    \label{eq:total_energy}
\end{multline}
where $C_{x}$ and $C_{xx}$ denote the set of pairs of states that are coupled by the driver and the catalyst respectively. We take $J_{xx}$ to be positive which is the case on which this paper is focused. With respect to the first term in the expression, which arises from the driver, energy reductions to $\braket{\hat{H}(s)}$ occur when the GS overlaps of pairs of states in $C_{x}$ have the same sign. For the second term, which arises as a result of the catalyst, pairs of states appearing in $C_{xx}$ will reduce the energy when the corresponding overlaps have the \textit{opposite} signs to each other. The final term in this expression, which arises as a result of the problem Hamiltonian, is irrelevant to our discussion since it is unaffected by the signs of the vector components.

Let us consider what this means for the relative signs of the instantaneous GS vector over the course of the anneal. At $s=0$, when the catalyst contribution is zero, it is energetically favourable for all the GS vector components to have the same sign. (This is consistent with the Perron–Frobenius theorem \cite{PFtheory}.) However for some $s$, given $J_{xx}$ is sufficiently large, we may expect one or more of the terms in the catalyst sum to become dominant enough that the energy would be lowered by certain components becoming negative. Understanding what combination of negative components will minimise the energy at any given $s$ is, in general, likely to be very complicated. We can simplify the situation somewhat by temporarily limiting ourselves to a single $XX-$coupling such that there are no conflicting pairs of coupled states within $C_{xx}$. In this case, it would be energetically favourable (as far as the catalyst term in equation \ref{eq:total_energy} is concerned) for the components of the instantaneous GS vector to split into two oppositely signed sets with one problem state from each pair in $C_{xx}$ appearing in each set. 

When considering how we expect this to impact how the signs to evolve over the course of the anneal, we might expect some of the pairs entering the sum to have a larger product such that they take opposite signs first, while it remains energetically favourable for others to keep the same sign as each other. The catalysts we consider in this work are those containing $XX-$terms chosen to couple some low energy local optimum, $\ket{E_a}$, participating in a PC with the ground state, to another low energy state, $\ket{E_{N_{xx}(a)}}$. By the fact that there is a crossing between $\ket{E_a(\lambda)}$ and $\ket{E_0(\lambda)}$ (\textit{i.e.}, the perturbed state, $\ket{E_a(\lambda)}$, becomes the instantaneous GS for some $s<s_\text{x}$) we can expect both $\ket{E_a}$ and $\ket{E_{N_{xx}(a)}}$ to have a large contribution to the instantaneous GS for some values of $s< s_\text{x}$. As such, we might expect this to be the first term in the catalyst sum to affect the signs of the problem state overlaps of the GS and so for the corresponding GS components, $\braket{E_0(s)|E_a}$ and $\braket{E_0(s)|E_{N_{xx}(a)}}$, to take opposite signs.

\begin{figure}[]
    \centering
    \includegraphics[scale=0.34]{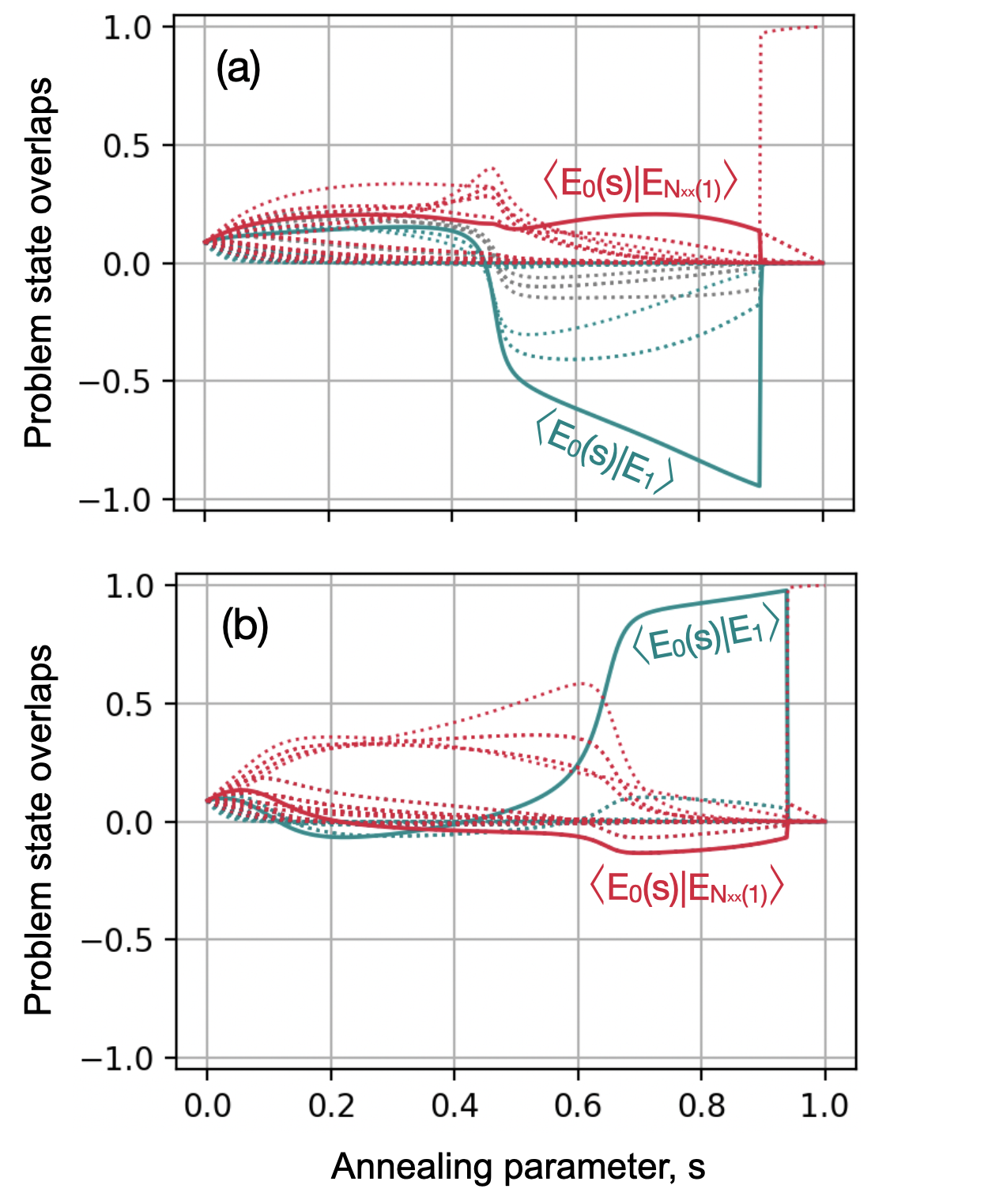}
    \caption{ Plots showing the evolution of the instantaneous GS associated with the anneal of an MWIS instance with $n_0=3$, $n_1=4$, $W_0=1.05$, $W_1=1.00$ and $J_{zz}=5$. Plot (a) shows the evolution in the presence of a catalyst containing a single coupling within  $G_1$, introduced with $J_{xx}=2$. Plot (b) shows the evolution when the catalyst contains all-to-all (six) couplings within $G_1$, all introduced with $J_{xx}=2$. In each case, the GS overlaps with $\ket{E_1}$ and $\ket{E_{N_{xx}(1)}}$ are plotted with solid teal and pink curves respectively. The other states are shown with dashed lines and are colour coded by which state they are closest to in Hamming distance. States that are equidistant from the two states are shown in grey. }
\label{fig:sign_changes}
\end{figure}

Looking at the signs of the GS coefficients for various example graphs we find that this is indeed the case -- with the rest of the overlaps sharing the sign of whichever state, $\ket{E_a}$ or $\ket{E_{N_{xx}(a)}}$, they are closer to, suggesting that the effect of the terms in the driver sum in equation \ref{eq:total_energy} is dominating over the other catalyst terms. An example is presented in figure \ref{fig:sign_changes}(a). The precise problem parameters are not of interest here but can be found in the figure caption. Crucially, there is a PC between the GS and the excited state associated with $G_1$, such that we are motivated to include an $XX-$coupling within $G_1$. The overlaps of the instantaneous GS with $\ket{E_1}$ and $\ket{E_{N_{xx}(1)}}$ are plotted in solid teal and pink respectively. The overlaps with the states that are closer in Hamming distance to each are plotted in the corresponding colours with dashed lines. States that are equidistant from the two states are shown in grey. We see that, at $s \approx 0.45$, the GS overlaps with $\ket{E_1}$ and $\ket{E_{N_{xx}(1)}}$ take different signs -- with $\braket{E_0(s)|E_1}$ becoming negative. Furthermore, we see that the overlaps with the states that are closer in Hamming distance to $\ket{E_1}$ also become negative.

We do not necessarily expect this exact pattern to hold for all settings where a single $XX-$term is introduced. The crucial observation however is as follows: In this work, we have motivated the introduction of $XX-$terms that couple problem states together that have a large presence in the perturbed problem state, $\ket{E_a(\lambda)}$, responsible for a PC with the GS. The purpose of these couplings is to introduce large terms into the sum in equation \ref{eq:non-stoq-pert} so that $\ket{E_a(\lambda)}$ receives large positive perturbations. Such couplings will only result in a \textit{positive} perturbation if the overlaps with the pairs of coupled problem states share the same sign. The preceding discussion however suggests that these overlaps will take opposite signs for sufficiently large $J_{xx}$ precisely \textit{because} of their large contribution to $\ket{E_a(\lambda)}$. At this point equation \ref{eq:non-stoq-pert} implies that continuing to increase $J_{xx}$ will result in a \textit{negative} perturbation to $\ket{E_a(\lambda)}$. This suggests a limit to the gap enhancement that can be achieved in this way before we start to see it begin to close again -- which is in line with our numerical results.

Let us now consider how this situation is altered by more than one $XX-$coupling being introduced into the catalyst. Figure \ref{fig:sign_changes}(b) corresponds to the case where the catalyst contains all-to-all $XX-$couplings within $G_1$ such that six terms are included in total. The overlaps with the different problem states are coloured as in (a). In this case however, rather than having a single state, $\ket{E_{N_{xx}(1)}}$, that the catalyst couples $\ket{E_1}$ to, we now have six states that are all symmetrically equivalent under the total Hamiltonian. These are the six states corresponding to sets consisting of all but two of the vertices in $G_1$. By virtue of being symmetrically equivalent, their overlaps with the GS have the same evolution -- shown by the solid pink curve. For values of $s$ at which we see negative overlaps, we observe that the overlap of the instantaneous ground state with $\ket{E_1}$ does \textit{not} consistently have the opposite sign to the states it becomes coupled to by the catalyst and that the other states do not group with regard to sign as they did in the case with a single $XX-$coupling. We also see that the magnitude of the negative overlaps are suppressed compared to the case where a single coupling is used. 

We suggest that these findings relate to the fact that the catalyst now contains conflicting pairs of states in $C_{xx}$ with non-negligible contributions to the GS. That is, $\ket{E_1}$ is coupled to each state corresponding to a set that consists of all but two vertices in $G_1$ and these states also become coupled to each other. It is clear that not all of the pairs of states can have overlaps with the opposite sign to each other. While this particular observation relates specifically to our graph structure and choice of catalyst, we suggest that introducing more $XX-$couplings will tend to result in more conflicting pairs of problem states such that these findings are likely to extend to other settings as well. This suggests that using multiple couplings will tend to suppress the problem of key pairs of overlaps taking the opposite sign and so result in the capacity for greater gap enhancement. This is again in agreement with the numerical results in section \ref{sec:numerics}.

These insights also suggest that it may be possible to further limit the appearance of negative vector components, and thereby reduce the precision needed in $J_{xx}$, by more careful design of the catalyst graph.

\section{ Instance construction \label{app:instance_construction} }

This appendix describes our process for generating the instances used in section \ref{sec:algorithm}. As discussed in section \ref{sec:algorithm_instances}, the basic procedure is to randomly generate the properties of a set of local-optima and then to seed these within an $18-$spin graph. 

First the properties of the maximally independent-sets that form the global and local optima are generated as follows:
\begin{itemize}
    \item Select the number of local optima to be generated, $3 \leq N_l \leq 7$.
    \item Allocate a total weight to the one global and $N_l$ local optima. The global optimum is given total weight of $W_0 = 2.05$. The local optima are randomly assigned weights, $W_a$ ($1 \leq a \leq N_l$), between $0.9$ and $2.0$.
    \item Select the number of vertices in the global optimum, $n_0$, between $2$ and $4$. 
    \item Select the number of vertices in each of the local optima, $n_a$ ($1 \leq a \leq N_l$), between $n_0+2$ and $n_0+5$. 
\end{itemize}
By having $n_a > n_0$ $\forall$ $a>0$, we ensure that the excited states associated with the local optima are likely to create PCs with the ground state. The decision to have the total weights of the local optima, $W_a$, separated from the total global optimum weight, $W_0=2.05$, by a minimum of $0.05$ is to ensure that there is an appreciable energy separation between the ground and first excited state in the problem spectrum. The reasoning behind this is that we are interested in whether or not the catalyst-construction process is able to remove the PCs present in the spectrum. Thus, we want the gap minimum of the spectra without PCs to be large enough for a significant final GS fidelity to be achieved such that this can be used as a success metric. 

\begin{figure}[]
    \centering
    \includegraphics[scale=0.28]{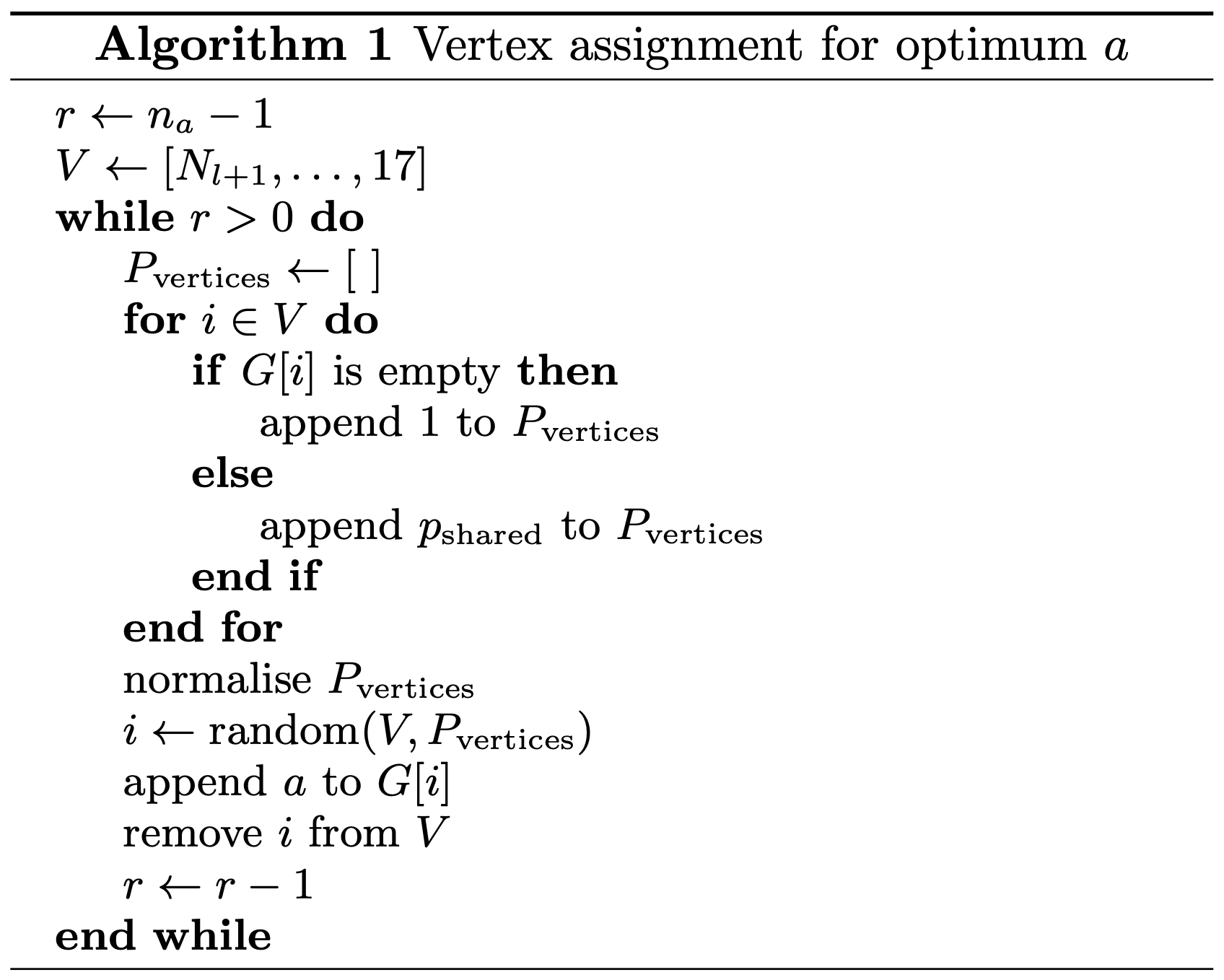}
\label{alg:vertex-assigment}
\end{figure}

Once the properties of the global and local optima have been generated, we want to seed them within an 18-vertex graph. To do this, we represent the graph as dictionary where the keys are the vertices of the graph and the entries are lists of indices indicating the optima that the vertex is associated with. For instance $G[6]=[0,2]$ indicates that vertex $6$ appears in the global optimum of the problem as well as the local optimum with index $2$.

We initialise the dictionary as 
\begin{equation}
    G = \{ 0:[0], 1:[1], \dots , N_l:[N_l], N_{l+1}:[\text{ }] , \dots , 17:[\text{ }] \}
\end{equation}
such that each optimum starts off with one vertex assigned to it. We then go through the global and each of the $N_l$ local optima in turn, randomly selecting vertices until they have each been assigned the correct number of vertices, $n_a$. The process of doing this for a given optimum, $0 \leq a \leq N_l$, is shown in algorithm $1$. The potential pool of vertices, $V$, that we select from does not include the first $N_l+1$ vertices such that each of these vertices remain allocated to one optimum alone. This ensures that the different optima remain distinct. We introduce a parameter, $p_\text{shared}$, which controls the extent to which we suppress the probability of selecting a vertex that already has already been assigned to an optimum. For the problem set generated for use in this work, $p_\text{shared}=0.5$. 

It is now possible to introduce the edges that will result in the intended maximally independent sets. This is done by introducing an edge between every pair of vertices that have no set allocations in common. For instance, two vertices $i$ and $j$ would have an edge between them if $G[i]=[1,3]$ and $G[j]=[0,4,5]$ but would not if $G[i]=[1,3]$ and $G[j]=[0,3]$. 

Finally, we must allocate weights to the vertices such that the different maximally independent sets defined at the start have the correct total weights. This is done according algorithm $2$ which cycles through each of the $N_l+1$ optima in turn (\textit{i.e.}, the one global and $N_l$ local optima). In this process we use the list $W_\text{vertices}$ to track the weights assigned to each vertex. Until a weight has been assigned to a vertex, its weight is listed as $0$. We use $W_\text{optima}$ to track the weight accumulated by the global and local optima as the algorithm progresses. The values in this list are re-calculated after each update to the vertex weights.

\begin{figure}[]
    \centering
    \includegraphics[scale=0.28]{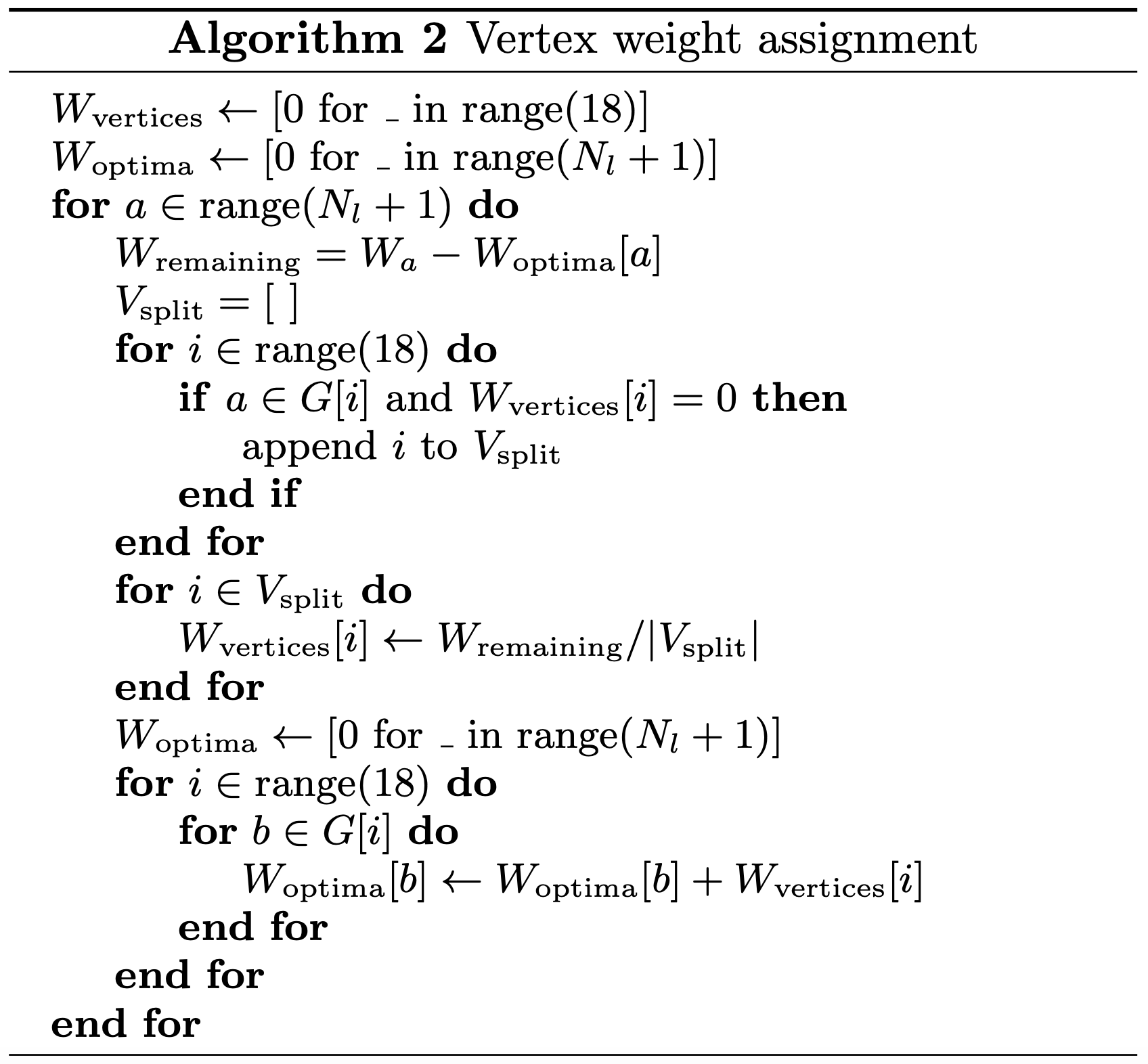}
\label{alg:weight-assigment}
\end{figure}

\end{document}